\documentclass[a4paper,10pt]{article}

\usepackage[utf8]{inputenc}
\usepackage[spanish,english]{babel}
\usepackage{graphicx}
\usepackage{amssymb}
\usepackage{amsmath}
\usepackage{hyperref}
\usepackage{color}
\usepackage{latexsym,dsfont,mathtools}
\usepackage{theorem,titletoc,titlesec}
\usepackage[normalem]{ulem}
\usepackage{enumitem}
\usepackage{tikz}
\usepackage{tabularx}
\usetikzlibrary{matrix}
\usepackage{adjustbox}
\usepackage{soul}
\newtheorem{theorem}{Theorem}[section]
\newtheorem{corollary}{Corollary}[section]
\newtheorem{proposition}{Proposition}[section]
\newtheorem{lemma}{Lemma}[section]
\newtheorem{example}{Example}[section]

\newenvironment{proof}[1][Proof.]{\vspace{0.5em}\textbf{#1} }{\
\hfill\rule{0.5em}{0.5em}}

\usepackage{multirow}
\usepackage{tabularx}
\usepackage{xcolor}

\newcommand{\Z}{\mathbb{Z}}

\newcommand{\zero}{{\mathbf{0}}}
\newcommand{\one}{{\mathbf{1}}}

\newcommand{\C}{{\cal C}}

\newcommand{\wt}{{\rm wt}}

\newcommand{\two}{{\mathbf{2}}}
\newcommand{\four}{{\mathbf{4}}}

\newcommand{\uu}{\mathbf{u}}
\newcommand{\vv}{\mathbf{v}}
\newcommand{\ww}{\mathbf{w}}
\newcommand{\zz}{\mathbf{z}}
\newcommand{\rank}{\operatorname{rank}}
\newcommand{\kernel}{\operatorname{ker}}

\newcommand{\cA}{{\cal A}}
\newcommand{\cH}{\cal{H}}

\begin{document}

\title{Linearity and Classification of $\mathbb{Z}_2\mathbb{Z}_4\mathbb{Z}_8$-Linear Hadamard Codes\thanks{This work has been partially supported by the Spanish MICIN under Grant PID2022-137924NB-I00 (AEI /10.13039/501100011033), by the Catalan AGAUR scholarship 2020 FI SDUR 00475 and grant 2021 SGR 00643.
}\thanks{the material in this paper was presented in part at the IEEE International Symposium on Information Theory (ISIT 2023), 25-30 June 2023  \cite{ISIT2023}.}
}

\author{Dipak K. Bhunia, Cristina Fern\'andez-C\'ordoba, Merc\`e Villanueva}
\date{
\footnotesize Department of Information and Communications
Engineering,\\
\footnotesize Universitat Aut\`{o}noma de Barcelona,\\
\footnotesize 08193 Cerdanyola del Vall\`{e}s, Spain\\
\footnotesize E-mails: \{Dipak.Bhunia, Cristina.Fernandez, Merce.Villanueva\}@uab.cat
}
 % \IEEEauthorblockA{Universitat Aut\`{o}noma de Barcelona \\ 
 %                   08193 Cerdanyola del Vall\`{e}s, Spain \\
 %                   Email: \{cristina.fernandez, carlos.vela, merce.villanueva\}@uab.cat}


\maketitle

%%%%%%
%% Abstract: 
%% If your paper is eligible for the student paper award, please add
%% the comment "THIS PAPER IS ELIGIBLE FOR THE STUDENT PAPER
%% AWARD." as a first line in the abstract. 
%% For the final version of the accepted paper, please do not forget
%% to remove this comment!
%%
\begin{abstract}
 The $\mathbb{Z}_2\mathbb{Z}_4\mathbb{Z}_8$-additive codes are subgroups of $\mathbb{Z}_2^{\alpha_1} \times \mathbb{Z}_4^{\alpha_2} \times \mathbb{Z}_8^{\alpha_3}$. A $\mathbb{Z}_2\mathbb{Z}_4\mathbb{Z}_8$-linear  Hadamard code is a Hadamard code which is the Gray map image of a $\mathbb{Z}_2\mathbb{Z}_4\mathbb{Z}_8$-additive code.  A recursive construction of $\mathbb{Z}_2\mathbb{Z}_4\mathbb{Z}_8$-additive Hadamard codes of type $(\alpha_1,\alpha_2, \alpha_3;t_1,t_2, t_3)$ with $\alpha_1 \neq 0$, $\alpha_2 \neq 0$, $\alpha_3 \neq 0$,  $t_1\geq 1$, $t_2 \geq 0$, and $t_3\geq 1$  is known.
In this paper, we generalize some known results for $\mathbb{Z}_2\mathbb{Z}_4$-linear Hadamard codes to $\mathbb{Z}_2\mathbb{Z}_4\mathbb{Z}_8$-linear Hadamard codes with $\alpha_1 \neq 0$, $\alpha_2 \neq 0$, and $\alpha_3 \neq 0$. 
First, we show for which types the corresponding $\mathbb{Z}_2\mathbb{Z}_4\mathbb{Z}_8$-linear Hadamard codes of length $2^t$ are nonlinear. For these codes, we compute the kernel and its dimension, which allows us to give a partial classification of these codes. 
Moreover, for $3 \leq t \leq 11$, we give a complete classification by providing the exact amount of nonequivalent such codes. We also prove the existence of several families of infinite such nonlinear $\mathbb{Z}_2\mathbb{Z}_4\mathbb{Z}_8$-linear Hadamard codes, which are not equivalent to any other constructed $\mathbb{Z}_2\mathbb{Z}_4\mathbb{Z}_8$-linear Hadamard code, nor to any $\mathbb{Z}_2\mathbb{Z}_4$-linear Hadamard code, nor to any previously constructed $\mathbb{Z}_{2^s}$-linear Hadamard code with $s\geq 2$, with the same length $2^t$.

\end{abstract}

%\end{titlepage}

%\newpage
%\tableofcontents
%\newpage

\section{Introduction}
Let $\Z_{2^s}$ be the ring of integers modulo $2^s$ with $s\geq1$. The set of
$n$-tuples over $\Z_{2^s}$ is denoted by $\Z_{2^s}^n$. In this paper,
the elements of $\Z^n_{2^s}$ will also be called vectors. 
A code over $\Z_2$ of length $n$ is a nonempty subset of $\Z_2^n$,
and it is linear if it is a subspace of $\Z_{2}^n$. Similarly, a nonempty
subset of $\Z_{2^s}^n$ is a $\Z_{2^s}$-additive code if it is a subgroup of $\Z_{2^s}^n$. A $\Z_2\Z_4\Z_8$-additive code is a subgroup of $\Z_2^{\alpha_1}  \times \Z_4^{\alpha_2} \times \Z_8^{\alpha_3}$. Note that a $\Z_2\Z_4\Z_8$-additive code is a linear code over $\Z_2$ when $\alpha_2=\alpha_3=0$,  a $\Z_4$-additive or $\Z_8$-additive code when $\alpha_1=\alpha_3=0$ or $\alpha_1=\alpha_2=0$, respectively, and a $\Z_2\Z_4$-additive code when $\alpha_3=0$. 
The order of a vector $u\in \Z_{2^s}^n$, denoted by $o(u)$, is the smallest positive integer $m$ such that $m u =(0,\dots,0)$. Also, the order of a vector $\mathbf u\in \Z_2^{\alpha_1}\times\Z_4^{\alpha_2} \times\Z_8^{\alpha_3}$, denoted by $o(\mathbf u)$, is the smallest positive integer $m$ such that $m \mathbf u =(0,\dots,0\mid 0,\dots,0 \mid 0,\dots,0)$.

Two binary codes $C_1$ and $C_2$ are said to be equivalent if there is a vector $\mathbf{a} \in \Z^n_2$ and a permutation of coordinates $\pi$ such that $C_2=\{\mathbf{a}+\pi(\mathbf{c}):\mathbf{a} \in C_1\}$.
The Hamming weight of a vector $u\in\Z_{2}^n$, denoted by $\wt_H(u)$, is
the number of nonzero coordinates of $u$. The Hamming distance of two
vectors $u,v\in\Z_{2}^n$, denoted by $d_H(u,v)$, is the number of
coordinates in which they differ.  Note that $d_H(u,v)=\wt_H(u-v)$. The minimum distance of a code $C$ over $\Z_2$ is $d(C)=\min \{ d_H(u,v) : u,v \in C, u \not = v  \}$.

In \cite{Sole}, a Gray map  from $\Z_4$ to $\Z_2^2$ is defined as
$\phi(0)=(0,0)$, $\phi(1)=(0,1)$, $\phi(2)=(1,1)$ and $\phi(3)=(1,0)$. There exist different generalizations of this Gray map, which go from $\Z_{2^s}$ to
$\Z_2^{2^{s-1}}$ \cite{Carlet,Codes2k,dougherty,Nechaev,Krotov:2007}.
The one given in \cite{Nechaev} can be defined in terms of the elements of a Hadamard code \cite{Krotov:2007}, and Carlet's Gray map \cite{Carlet} is a particular case of the one given in \cite{Krotov:2007} 
satisfying $\sum \lambda_i \phi(2^i) =\phi(\sum \lambda_i 2^i)$ \cite{KernelZ2s}. 
In this paper, we focus on Carlet's Gray map \cite{Carlet}, from $\Z_{2^s}$ to $\Z_2^{2^{s-1}}$, which is also a particular case of the one given in \cite{ShiKrotov2019}. Specifically,
\begin{gather}\label{eq:GrayMapCarlet}
\phi_s(u)=(u_{s-1},u_{s-1},\dots,u_{s-1})+(u_0,\dots,u_{s-2})Y_{s-1},
\end{gather}
where $u\in\Z_{2^s}$; $[u_0,u_1,\dots,u_{s-1}]_2$ is the binary expansion of $u$, that is, $u=\sum_{i=0}^{s-1}u_i2^i$ with $u_i\in \{0,1\}$; and $Y$ is a matrix of size $(s-1)\times 2^{s-1}$ whose columns are all the vectors in $\Z_2^{s-1}$. Without loss of generality, we assume that the columns of $Y_{s-1}$ are ordered in ascending order, by considering the elements of $\mathbb{Z}_2^{s-1}$ as the binary expansions of the elements of $\mathbb{Z}_{2^{s-1}}$. Note that $\phi_1$ is the identity map.
%and $(u_{s-1},\dots,u_{s-1})$. % and $(u_0,\dots,u_{s-2})Y_{s-1}$ are binary vectors of length $2^{s-1}$, and that the rows of $Y_{s-1}$ form a basis of a first order Reed-Muller code after adding the all-one row. 
We define $\Phi_s:\Z_{2^s}^n \rightarrow \Z_2^{n2^{s-1}}$ as the component-wise extended map of $\phi_s$.
We can also define a Gray map $\Phi$ from $\Z_2^{\alpha_1}  \times \Z_4^{\alpha_2} \times \Z_8^{\alpha_3}$ to $\Z_2^n$, where $n=\alpha_1+2\alpha_2+4\alpha_3$, as follows:
$$
\Phi(u_1\mid u_2 \mid u_3)=(u_1, \Phi_2(u_2),\Phi_3(u_3)),
$$
for any $u_i \in \Z_{2^i}^{\alpha_i}$, where $1\leq i\leq 3$.

Let $\C \subseteq \Z_{2^s}^n$ be a $\Z_{2^s}$-additive code of length $n$. We say that the Gray map image of $\C$, say
$C=\Phi_s(\C)$ is a $\Z_{2^s}$-linear code of length $n2^{s-1}$. 
Since $\C$ is a subgroup of
$\Z_{2^s}^n$, it is isomorphic to
$\Z_{2^s}^{t_1}\times\Z_{2^{s-1}}^{t_2}\times
\dots\times\Z_2^{t_s}$, and we say that $\C$, or equivalently
$C=\Phi_s(\C)$, is of type $(n;t_1,\dots,t_{s})$.
%Note that $|\C|=2^{st_1}2^{(s-1)t_2}\cdots 2^{t_s}$. 
Similarly, if $\C \subseteq \Z_2^{\alpha_1}  \times \Z_4^{\alpha_2} \times \Z_8^{\alpha_3}$
%{\color{blue}\xout{Let $\C$ be}} 
is a $\Z_2\Z_4\Z_8$-additive code,  we say that its Gray map image $C=\Phi(\C)$ is a $\Z_2\Z_4\Z_8$-linear code of length $\alpha_1+2\alpha_2+4\alpha_3$. Since $\C$ can be seen as a subgroup of
$\Z_8^{\alpha_1+\alpha_2+\alpha_3}$, it is isomorphic to
$\Z_8^{t_1}\times\Z_4^{t_2} \times\Z_2^{t_3}$, and we say that $\C$, or equivalently
$C=\Phi(\C)$, is of type $(\alpha_1, \alpha_2,\alpha_3;t_1,t_2,t_3)$.
Note that a $\Z_2\Z_4$-linear code $\C$ \cite{ccsg,BookZ2Z4} can be seen as a $\Z_2\Z_4\Z_8$-linear code of type $(\alpha_1,\alpha_2,0;0,t_2,t_3)$. In this case, we also say that the type of $\C$ is directly $(\alpha_1, \alpha_2;t_2,t_3)$. Unlike linear codes over finite fields, linear codes over rings do not have a basis, but there exists a generator matrix for these codes having a minimum number of rows. If $\C$ is a $\Z_2\Z_4\Z_8$-additive code of type $(\alpha_1, \alpha_2,\alpha_3;t_1,t_2,t_3)$, then $|\C|=8^{t_1}4^{t_2} 2^{t_3}$ and there exist a generator matrix with $t_1+t_2+t_3$ rows.

Two structural properties of codes over $\Z_2$ are the rank and
dimension of the kernel. The rank of a code $C$ over $\Z_2$ is simply the
dimension of the linear span, $\langle C \rangle$,  of $C$.
The kernel of a code $C$ over $\Z_2$ is defined as
$\mathrm{K}(C)=\{x\in \Z_2^n : x+C=C \}$ \cite{BGH83}. If the all-zero vector belongs to $C$,
then $\mathrm{K}(C)$ is a linear subcode of $C$.
Note also that if $C$ is linear, then $K(C)=C=\langle C \rangle$.
We denote the rank of $C$ as $\rank(C)$ and the dimension of the kernel as $\kernel(C)$.
These parameters can be used to distinguish between nonequivalent codes since equivalent ones have the same rank and dimension of the kernel.

A binary code of length $n$, $2n$ codewords and minimum distance $n/2$ is called a Hadamard code. Hadamard codes can be constructed from Hadamard matrices \cite{Key,WMcwill}.
Note that linear Hadamard codes are in fact first order Reed-Muller codes, or equivalently, the dual of extended Hamming codes \cite{WMcwill}. The $\Z_{2^s}$-additive codes such that after the Gray map $\Phi_s$ give
Hadamard codes are called $\Z_{2^s}$-additive Hadamard codes and the
corresponding images are called $\Z_{2^s}$-linear Hadamard codes. 
Similarly, the $\Z_2\Z_4\Z_8$-additive codes such that after the Gray map $\Phi$ give
Hadamard codes are called $\Z_2\Z_4\Z_8$-additive Hadamard codes and the
corresponding images are called $\Z_2\Z_4\Z_8$-linear Hadamard codes.

It is known that $\Z_4$-linear Hadamard codes (that is, $\Z_2\Z_4$-linear Hadamard codes with $\alpha_1 =0$) and $\Z_2\Z_4$-linear Hadamard codes with $\alpha_1\not = 0$ can be classified by using either the rank or the dimension of the kernel \cite{Kro:2001:Z4_Had_Perf,PRV06}. Moreover, in \cite{KV2015}, it is shown that each $\Z_4$-linear Hadamard code is  equivalent to a  $\Z_2\Z_4$-linear Hadamard code with $\alpha_1 \not =0$.
Later, in \cite{KernelZ2s,HadamardZps,EquivZ2s,ZpsEquivalance}, a recursive construction for $\Z_{p^s}$-linear Hadamard codes, with $p$ prime, is described, the linearity is established, and a partial classification by using the dimension of the kernel is obtained, giving the exact amount of nonequivalent such codes for some parameters. 
In \cite{fernandez2019mathbb}, a complete classification of $\Z_8$-linear Hadamard codes by using the rank and dimension of the kernel is provided, giving the exact amount of nonequivalent such codes.  For any $t\geq 2$, the full classification of $\Z_p\Z_{p^2}$-linear Hadamard codes of length $p^t$, with $\alpha_1\neq 0$, $\alpha_2\neq 0$, and $p\geq 3$ prime, is given in \cite{ZpZp2Construction,ZpZp2Classification}, by using just the dimension of the kernel. 

The paper contributes
to the study of codes over rings $\mathbb{Z}_{p^s}$, which were first studied by Blake \cite{blake} and
Shankar \cite{shankar} in 1975 and 1979, respectively. These codes have become more significant after the
publication of \cite{Sole}. It is also important to note that Hadamard codes are two weight codes, which have been widely studied in \cite{ShiTwoHomWeight,TwoWeightSole}. On the other hand, the classification of nonlinear  Hadamard codes is still an open problem. By giving an additive structure, as $\mathbb{Z}_{p^s}$-linear,  $\mathbb{Z}_p\mathbb{Z}_{p^2}$-linear or $\mathbb{Z}_2\mathbb{Z}_{4}\mathbb{Z}_8$-linear codes, to some of them, and showing whether they are equivalent or not among them, we are providing a partial classification for these codes.

 From a more practical point of view, since Hadamard codes are optimal and have a high correction capability, they appear in different aspects related to the transmission of information, such as in digital communication with satellites \cite{H07}, in CDMA phones to modulate the transmission of information and minimize interference with other transmissions \cite{SBT98} and, in general, in different OCDMA multiple access systems to allow access to multiple users asynchronously and simultaneously \cite{HYT04}. Other applications are found in cryptography \cite{Nyb91} or in information hiding (steganography and watermarking) \cite{YLL03}. See \cite{H07} for more applications in other fields.

 This paper is focused on $\Z_2\Z_4\Z_8$-linear Hadamard codes with $\alpha_1\not =0$, $\alpha_2\not =0$, and $\alpha_3\not =0$, generalizing some results given for $\Z_2\Z_4$-linear Hadamard codes with $\alpha_1\not =0$ and $\alpha_2\not =0$ in \cite{PRV06} related to the linearity, kernel, its dimension, and the classification of such codes. These codes are also compared with the $\Z_{2^s}$-linear Hadamard codes with $s\geq 2$ considered in \cite{KernelZ2s}.
This paper is organized as follows.
In Section \ref{Sec:GrayMap}, we recall some properties of the generalized Gray map considered in this paper, the recursive construction of  $\Z_2\Z_4\Z_8$-linear Hadamard codes of type $(\alpha_1,\alpha_2, \alpha_3;t_1,t_2, t_3)$, with $\alpha_1 \neq 0$, $\alpha_2 \neq 0$, $\alpha_3 \neq 0$, and some known results, given in \cite{Z2Z4Z8Construction}. %and $t_1\geq 1$, $t_2 \geq 0$, $t_3\geq 1$.
In Sections \ref{Sec:Linearity}  and \ref{Sec:Kernel}, we establish for which types these codes are linear, and
we give the kernel and its dimension whenever they are nonlinear. 
In Section \ref{Sec:classification}, we prove the existence of several families of infinite such nonlinear $\Z_2\Z_4\Z_8$-linear Hadamard codes, which are not equivalent to any other constructed $\Z_2\Z_4\Z_8$-linear Hadamard code, nor to any $\Z_2\Z_4$-linear Hadamard code, nor to any previously constructed $\Z_{2^s}$-linear Hadamard code with $s\geq 2$, with the same length $2^t$. We also give a complete classification of  $\Z_2\Z_4\Z_8$-linear Hadamard codes of length $2^t$ for $3 \leq t \leq 11$,  by providing the exact amount of nonequivalent such codes.
Finally, in Section \ref{sec:conclusions}, we give some conclusions and further research on this topic.

\section{Preliminary results}\label{Sec:GrayMap}

In this section, we first recall some properties of the generalized Gray map $\phi_s$. 
%give the definition of the generalized Gray maps considered in this paper for elements of $\Z_4$ and $\Z_8$, and in general of $\Z_{2^s}$, $s\geq 2$. We also include some of its properties used in the paper. 
Then, we also recall the recursive construction of $\Z_2\Z_4\Z_8$-additive Hadamard codes of type $(\alpha_1,\alpha_2, \alpha_3;t_1,t_2,t_3)$, with  $\alpha_1\not =0$, $\alpha_2\not =0$, and $\alpha_3\not =0$, and some known results, given in \cite{Z2Z4Z8Construction}.

\medskip
% We consider the Carlet's Gray map from $\Z_{2^s}$ to $\Z_2^{2^{s-1}}$ \cite{Carlet} given in (\ref{eq:GrayMapCarlet}).
% For $s=2$ and $s=3$, the Gray maps $\phi_2$ and $\phi_3$ considered in the paper for the elements of $\Z_4$ and $\Z_8$, respectively, are the following:
% \begin{equation*}
% \left.\begin{array}{cccccccccl}
% \phi_2: &\Z_4 \longrightarrow \Z_2^2 &&& \phi_3: & \Z_8 \longrightarrow \Z_2^4\\
% &0\mapsto(0,0) &&& & 0\mapsto (0,0,0,0) \\
% &1\mapsto(0,1)&&& &1\mapsto (0,1,0,1)\\
% &2\mapsto(1,1)&&& &2 \mapsto (0,0,1,1)\\
% &3\mapsto(1,0)&&& &3 \mapsto (0,1,1,0)\\
%  &            &&& &4 \mapsto (1,1,1,1)\\
%   &           &&& &5 \mapsto (1,0,1,0)\\
%    &          &&& &6 \mapsto (1,1,0,0)\\
%     &         &&& &7 \mapsto (1,0,0,1).\\
% \end{array}\right.
% \end{equation*}

Let $u',v'\in\Z_{2^s}$ and $[u'_0,u'_1,\dots,u'_{s-1}]_2$, $[v'_0,v'_1,\dots,v'_{s-1}]_2$ be the
binary expansions of $u'$ and $v'$, respectively, i.e. $u'=\sum_{i=0}^{s-1}u'_i2^i$ and $v'=\sum_{i=0}^{s-1}v'_i2^i$. We define the operation  ``$\odot$''  between elements  $u'$ and $v'$ in $\Z_{2^s}$ as $u'\odot v'=\sum_{i=0}^{s-1}\xi_i2^i$, where 
$$
\xi_i=\left\lbrace\begin{array}{ll}
 1 & \textrm{if } \quad u'_i+v'_i =2,\\
0 &  \textrm{otherwise}.\\
\end{array}\right.
$$
Note that  the binary expansion of $u'\odot v'$ is $[\xi_0,\xi_1,\dots,\xi_{s-1}]_2$, where $\xi_i \in \{0,1\}$. 
We denote in the same way, ``$\odot$'', the component-wise operation.
 For $\uu=(u_1\mid u_2  \mid u_3), \vv=(v_1\mid v_2\mid v_3) \in \Z_2^{\alpha_1} \times \Z_{4}^{\alpha_2} \times \Z_{8}^{\alpha_3}$, we denote $\uu \odot \vv=(u_1 \odot v_1 \mid u_2 \odot v_2 \mid u_3 \odot v_3)$. Note that $2(\uu \odot \vv)=(\zero \mid 2(u_2 \odot v_2)  \mid 2(u_3\odot v_3))$.
 
From \cite{HadamardZps}, we have the following results:

% \begin{lemma}\cite{HadamardZps}\label{lemma1}
% Let $u \in \Z_{2^s}$ and $\lambda \in \Z_2$. Then, 
% $\phi_s(u+\lambda 2^{s-1})= \phi_s(u)+ (\lambda, \lambda ,\dots,\lambda)$.
% \end{lemma}

\begin{corollary}\cite{HadamardZps}\label{coro4}
 Let $\lambda, \mu \in \Z_2$. Then,  $\phi_s(\lambda \mu 2^{s-1})= \lambda \phi_s(\mu 2^{s-1})=\lambda \mu \phi_s(2^{s-1})$.
\end{corollary}
% \begin{proposition}\label{prop1}
% Let $u, v$ $\in \Z_{2^s}$. Then, $\phi (u) + \phi (v) = \phi (u\oplus_p v)$.
% \end{proposition}

% \begin{proposition}\label{pprop}
% Let $u$, $v$ $\in \Z_{2^s}$. Then, $u\oplus_p v=u+v-p(u\odot v)$.
% \end{proposition}

\begin{corollary}\cite{HadamardZps}\label{coro1}
Let $u, v$ $\in \Z_{2^s}$. Then, $\phi_s (u) + \phi_s (v) = \phi_s (u+v-2(u\odot v))$.
\end{corollary}

\begin{corollary}\cite{HadamardZps} \label{lemma2}
Let $u, v$ $\in \Z_{2^s}$. Then, $\phi_s(2^{s-1}u+v)= \phi_s(2^{s-1}u)+\phi_s(v)$.
\end{corollary}

\begin{corollary}\label{coro2}
Let $u, v$ $\in \Z_4$ and $[u_0,u_1]_2$, $[v_0,v_1]_2$ be the
binary expansions of $u$ and $v$, respectively. Then, $\phi_2 (2(u\odot v))=(\xi_0,\xi_0)$,
 where $\xi_0= 1$ when $u_0+v_0 \geq 2$ and $\xi_0=0$ otherwise.
\end{corollary}

\begin{corollary}\label{coroNew1}
Let $u, v$ $\in \Z_8$ and $[u_0,u_1,u_2]_2$, $[v_0,v_1,v_2]_2$ be the
binary expansions of $u$ and $v$, respectively. Then, $\phi_3 (2(u\odot v))=\phi_3(2\xi_0+4\xi_1)
= \phi_3 (2\xi_0)+(\xi_1,\xi_1,\xi_1,\xi_1)$,
 where $\xi_i= 1$ when $u_i+v_i = 2$ and $0$ otherwise.
\end{corollary}

%\begin{proposition}\cite{Carlet,GrayIsometry,HadamardZps}\label{disweight}
%Let $u,v\in\Z_{2^s}$. Then, $$d_H(\phi_s(u),\phi_s(v))=\wt_H(\phi_s(u-v)).$$
%\end{proposition}
%
%By Proposition \ref{disweight}, the $\Z_2\Z_4\Z_8$-linear codes obtained from the Gray map $\Phi$ are distance invariant, that is, the Hamming weight distribution is invariant under translation by a codeword. Therefore, their minimum distance coincides with the minimum weight.

 Let $\zero, \one,\two,\ldots, \mathbf{7}$ be the vectors having the elements $0, 1, 2, \ldots, 7$  repeated in each coordinate, respectively. If $A$ is a generator matrix of a $\Z_2\Z_4\Z_8$-additive code, that is, a subgroup of $\Z_2^{\alpha_1} \times \Z_4^{\alpha_2} \times \Z_8^{\alpha_3}$ for some integers $\alpha_1,\alpha_2,\alpha_3\geq 0$, then we denote by $A_1$ the submatrix of $A$ with the first $\alpha_1$ columns over $\Z_2$, $A_2$ the submatrix with the next $\alpha_2$ columns over $\Z_4$, and $A_3$ the submatrix with the last $\alpha_3$ columns over $\Z_8$. We have that $A=(A_1\mid A_2\mid A_3)$, where the number of columns of $A_i$ is $\alpha_i$ for $i\in \{1,2,3\}$.

Let $t_1\geq 1$, $t_2\geq 0$, and $t_3\geq 1$ be integers. Now, we construct recursively matrices $A^{t_1,t_2,t_3}$ having $t_1$ rows of order $8$, $t_2$ rows of order $4$, and $t_3$ rows of order $2$ as follows. First, we consider the following matrix:
\begin{equation}\label{eq:recGenMatrix0}
A^{1,0,1}=
\left(\begin{array}{cc|c|c}
1 & 1  & 2  &4 \\
0  & 1 &1  &1  \\
\end{array}\right).
\end{equation}
Then, we apply the following constructions. 
If we have a matrix  $A^{\ell-1,0,1}=(A_1 \mid A_2 \mid A_3)$, with $\ell \geq 2$, we may construct the matrix
\begin{equation}\label{eq:recGenMatrix1}
\footnotesize
A^{\ell,0,1}=
\left(\begin{array}{cc|ccccc|ccccc}
A_1 & A_1 &M_1 &A_2 &A_2 &A_2 &A_2 &M_2 &A_3 &A_3 &\cdots &A_3 \\
\mathbf{0}  & \mathbf{1} & \mathbf{1}  &\mathbf{0} &\mathbf{1}  &\mathbf{2} &\mathbf{3} &\mathbf{1} &\mathbf{0} &\mathbf{1} &\cdots &\mathbf{7} \\
\end{array}\right),
\end{equation}
where $M_1=\{\mathbf{z}^T:\mathbf{z}\in\lbrace2\rbrace\times \lbrace0,2\rbrace^{\ell-1}\}$ and $M_2=\{\mathbf{z}^T: \mathbf{z}\in\lbrace4\rbrace\times \lbrace0,2,4,6\rbrace^{\ell-1}\}$. We perform construction (\ref{eq:recGenMatrix1}) until $\ell=t_1$. If we have a matrix $A^{t_1,\ell-1,1}=(A_1\mid A_2 \mid A_3)$, with $t_1 \geq 1$ and $\ell\geq 1$, we may construct the matrix 
\begin{equation}\label{eq:recGenMatrix2}
\footnotesize
A^{t_1,\ell,1}=
\left(\begin{array}{cc|ccccc|cccc}
A_1 & A_1 &M_1 &A_2 &A_2 &A_2 &A_2 &A_3 &A_3 &A_3 &A_3 \\
\mathbf{0}  & \mathbf{1} & \mathbf{1}  &\mathbf{0} &\mathbf{1}  &\mathbf{2} &\mathbf{3}  &\mathbf{0} &\mathbf{2} &\mathbf{4} &\mathbf{6} \\
\end{array}\right),
\end{equation}
where $M_1=\{\mathbf{z}^T:\mathbf{z}\in\lbrace2\rbrace\times \lbrace0,2\rbrace^{t_1+\ell-1}\}$. We repeat construction (\ref{eq:recGenMatrix2}) until $\ell=t_2$. Finally, if we have a matrix $A^{t_1,t_2,\ell-1}=(A_1\mid A_2 \mid A_3)$, with $t_1\geq 1$, $t_2\geq 0$, and $\ell\geq 2$, we may construct the matrix
\begin{equation}\label{eq:recGenMatrix3}
A^{t_1,t_2,\ell}=
\left(\begin{array}{cc|cc|cc}
A_1 & A_1 & A_2 & A_2 &A_3 &A_3 \\
\mathbf{0}  & \mathbf{1} & \mathbf{0} &\mathbf{2}  &\mathbf{0}  &\mathbf{4}  \\
\end{array}\right).
\end{equation}
We repeat construction (\ref{eq:recGenMatrix3}) until $\ell=t_3$. Thus, in this way, we obtain $A^{t_1,t_2,t_3}$.

Summarizing, in order to achieve $A^{t_1,t_2,t_3}$ from $A^{1,0,1}$, first we add $t_1-1$ rows of order $8$  by applying construction (\ref{eq:recGenMatrix1}) $t_1-1$ times, starting from $A^{1,0,1}$ up to obtain $A^{t_1,0,1}$; then we add $t_2$ rows of order $4$ by applying construction (\ref{eq:recGenMatrix2}) $t_2$ times, up to generate $A^{t_1,t_2,1}$; and,  finally, we add $t_3-1$ rows of order $2$ by applying construction (\ref{eq:recGenMatrix3}) $t_3-1$ times to achieve $A^{t_1,t_2,t_3}$. Note that in the first row there is always the row $(\one \mid \two \mid \four)$.

\begin{example} \label{ex:MatricesA}
By using the constructions described in (\ref{eq:recGenMatrix1}), (\ref{eq:recGenMatrix2}), and (\ref{eq:recGenMatrix3}), we obtain the following matrices $A^{2,0,1}$, $A^{1,1,1}$ and $A^{1,1,2}$, respectively, starting from $A^{1,0,1}$ given in (\ref{eq:recGenMatrix0}):
\begin{equation}\label{eq:A201}
A^{2,0,1}=
\left(\begin{array}{cc|cc|cc}
11 &11 &22 &2 2 2 2 &4 4 4 4 &4 4 4 4 4 4 4 4 \\
0  1 &0  1 &0 2 &1 1 1 1 &0 2 4 6 &1 1 1 1 1 1 1 1 \\
0 0 &1 1 &1 1 &0 1 2 3 &1 1 1 1 &0 1 2 3 4 5 6 7\\ 
\end{array}\right),
\end{equation}
\begin{equation}\label{eq:A111}
A^{1,1,1}=
\left(\begin{array}{cc|cc|c}
 11 &11 &22&2222 &4444\\
 01&01 &02&1111 &1111\\
 00&11 &11&0123 &0246\\ 
\end{array}\right), %\mbox{and}
\end{equation}
$$
A^{1,1,2}=
\left(\begin{array}{cc|cc|cc}
 1111 & 1111 &222222 &222222 &4444 &4444\\
 0101 & 0101 &021111 &021111 &1111 &1111\\
 0011 & 0011 &110123 &110123 &0246 &0246\\ 
 0000 &1111  &000000 &222222 &0000 &4444\\
\end{array}\right).
$$
%In order to obtain $A^{2,1,1}$, we start with $A^{1,0,1}$,  we apply construction (\ref{eq:recGenMatrix1}) to obtain $A^{2,0,1}=(A_1\mid A_2\mid A_3)$ given in (\ref{eq:A201}), and then we apply (\ref{eq:recGenMatrix2}) to obtain 
% \begin{equation*}\label{eq:A_211}
%A^{2,1,1}=\left(\begin{array}{cc|@{}cccccc|cccc}
%      A_1  &A_1 &&\begin{matrix}
%            22 2 2\\
%            0 0 2 2\\
%            0 2 0 2\\
%      \end{matrix} &A_2  &A_2  &A_2  &A_2 &A_3 &A_3 &A_3 &A_3\\
%      \mathbf{0} &\mathbf{1} &&\mathbf{1} &\mathbf{0} &\mathbf{1} &\mathbf{2} &\mathbf{3} &\mathbf{0} &\mathbf{2} &\mathbf{4} &\mathbf{6} \\
%   \end{array}\right).
%\end{equation*}
\end{example}

\medskip
The $\Z_2\Z_4\Z_8$-additive code generated by $A^{t_1,t_2,t_3}$ is denoted by ${\cH}^{t_1,t_2,t_3}$, and the corresponding $\Z_2\Z_4\Z_8$-linear code $\Phi( {\cH}^{t_1,t_2,t_3})$ by $H^{t_1,t_2, t_3}$.

\begin{proposition}\cite{Z2Z4Z8Construction}\label{relation:t1a1a3}
Let $t_1\geq 1$, $t_2\geq 0$, and $t_3\geq 1$ be integers. Let ${\cH}^{t_1,t_2,t_3}$ be the $\Z_2\Z_4 \Z_8$-additive code of type $(\alpha_1, \alpha_2, \alpha_3; t_1,t_2,t_3)$ generated by $A^{t_1,t_2,t_3}$. Then,  
\begin{align*} \label{eq:propt1t2t3}
\begin{split}
    &\alpha_1=2^{t_1+t_2+t_3-1},\\ 
    &\alpha_1+2\alpha_2=4^{t_1+t_2}2^{t_3-1},\\ 
    &\alpha_1+2\alpha_2+ 4\alpha_3= 8^{t_1}4^{t_2}2^{t_3-1}.
\end{split}
\end{align*}
\end{proposition}

 \begin{theorem}\cite{Z2Z4Z8Construction}\label{Th:H3}
Let $t_1\geq 1$, $t_2\geq 0$, and $t_3\geq 1$ be integers. Then, the $\Z_2\Z_4 \Z_8$-additive code 
${\cH}^{t_1,t_2,t_3}$, generated by $A^{t_1,t_2,t_3}$, is a $\Z_2\Z_4 \Z_8$-additive Hadamard code of length $2^t$, with $t+1=3t_1+2t_2+t_3$.
\end{theorem}

\begin{example}\label{ex:linear101}
The $\Z_2\Z_4 \Z_8$-additive code ${\cH}^{1,0,1}$ generated by $A^{1,0,1}$, given in (\ref{eq:recGenMatrix0}),
is a $\Z_2\Z_{4}\Z_8$-additive Hadamard code of type $(2,1,1;1,0,1)$.  
We can write ${\cH}^{1,0,1}=\bigcup_{\alpha \in\{0,1\}} ({\cA} + \alpha (1,1\mid 2\mid 4))$, where ${\cA}=\{\lambda (0, 1\mid 1\mid 1): \lambda \in \Z_8\}$.
Thus, $H^{1,0,1}=\Phi({\cH}^{1,0,1})= \bigcup_{\alpha \in\{0,1\}} (\Phi({\cA}) + \alpha \textbf{1})$, where $\Phi({\cA})$ consists of all the rows of the Hadamard matrix 
\begin{equation*}
    H(2,4)=
    \left(\begin{array}{cccccccc}
0 &0 &0 &0 &0 &0 &0 &0  \\
0 &1 &0 &1 &0 &1 &0 &1   \\
0 &0 &1 &1 &0 &0 &1 &1  \\
0 &1 &1 &0 &0 &1 &1 &0 \\
0 &0 &0 &0 &1 &1 &1 &1 \\
0 &1 &0 &1 &1 &0 &1 &0 \\
0 &0 &1 &1 &1 &1 &0 &0 \\
0 &1 &1 &0 &1 &0 &0 &1 \\
\end{array}\right).
\end{equation*}
Note that $\Phi({\cA})$ is linear and the minimum distance of $\Phi({\cA})$ is $4$, so $H^{1,0,1}$ is a binary linear Hadamard code of length $8$. 
\end{example}

\section{Linearity of $\Z_2\Z_4\Z_8$-linear Hadamard codes}\label{Sec:Linearity}

The linearity of $\Z_4$-linear Hadamard codes was studied in \cite{Kro:2001:Z4_Had_Perf,PRV06}. In general, for $\Z_{p^s}$-linear Hadamard codes with $s\geq 2$ and $p$ prime, the results on the linearity are given in \cite{KernelZ2s} when $p=2$ and in \cite{HadamardZps} when $p\geq 3$. In \cite{PRV06}, it is shown that the $\Z_2\Z_4$-linear Hadamard codes of type $(\alpha_1,\alpha_2;1,t_3)$, or equivalently, the $\Z_2\Z_4\Z_8$-linear Hadamard codes of type $(\alpha_1, \alpha_2,0;0,1,t_3)$, with $\alpha_1\not =0$, $\alpha_2\not =0$, and $t_2\geq 1$, are the only ones which are linear when $\alpha_1\not =0, \alpha_2\not =0$, and $\alpha_3=0$. The linearity of $\Z_p\Z_{p^2}$-linear Hadamard codes, with $\alpha_1\not =0, \alpha_2\not =0$, and $p\geq 3$ prime, was studied in \cite{ZpZp2Construction}. The next results show that the Hadamard codes $H^{1,0,t_3}=\Phi({\cH}^{1,0,t_3})$, where $t_3 \geq 1$, are the only $\Z_2\Z_4\Z_8$-linear Hadamard codes $H^{t_1,t_2,t_3}$ %\cite{Z2Z4Z8Construction} 
with $\alpha_1 \not =0$, $\alpha_2\not =0$, and $\alpha_3\not =0$ which are linear.

\begin{proposition}
\label{Th:linear}
The $\Z_2\Z_4\Z_8$-linear Hadamard codes $H^{1,0,t_3}$, with $t_3 \geq 1$, are linear.
\end{proposition}
\begin{proof}
We prove this theorem by induction on $t_3 \geq 1$. By
Example \ref{ex:linear101},
$H^{1,0,1}$ is linear. Assume that $H^{1,0,t_3}=\Phi({\cH}^{1,0,t_3})$, with $t_3\geq 1$, is  linear. Let  ${\cH}={\cH}^{1,0,t_3+1}$ and $H=\Phi(\cH)$. Now, we have to show that $H$ is also linear. Let ${\cH}^{1,0,t_3}=({\cH}_1\mid {\cH}_2\mid {\cH}_3)$. By construction (\ref{eq:recGenMatrix3}),
\begin{align*}
    {\cH}=\bigcup_{\lambda \in  \{0,1\}}(({\cH}_1, {\cH}_1\mid {\cH}_2, {\cH}_2\mid {\cH}_3, {\cH}_3)+\lambda(\zero, \one \mid \zero, \two \mid \zero, \four)).
\end{align*}
By Corollaries \ref{coro4} and  \ref{lemma2}, 
\begin{align*}\label{eq:Ht3p1}
    H&=\bigcup_{\lambda \in \{0,1\}}(\Phi({\cH}_1, {\cH}_1\mid {\cH}_2, {\cH}_2\mid {\cH}_3, {\cH}_3)+\lambda(\zero, \one,  \zero, \one,  \zero, \one))\nonumber\\
    &=A_0 \cup A_1,
\end{align*}
where $A_\lambda=\Phi({\cH}_1, {\cH}_1\mid {\cH}_2, {\cH}_2\mid {\cH}_3, {\cH}_3)+\lambda(\zero, \one,  \zero, \one,  \zero, \one)$, $\lambda \in \{0,1\}$. Any element in $A_\lambda$ is of the form 
\begin{equation}\label{eq:Alambda}
\Phi(u_1, u_1\mid u_2, u_2 \mid u_3, u_3)+\lambda(\zero, \one,  \zero, \one,  \zero, \one),
\end{equation} for $(u_1\mid u_2 \mid u_3) \in ({\cH}_1\mid {\cH}_2\mid {\cH}_3)$. Note that (\ref{eq:Alambda}) is equal to 
$(u_1, u_1+\lambda \cdot \one, \Phi_2(u_2),\Phi_2(u_2)+\lambda \cdot \one, \Phi_3(u_3),\Phi_3(u_3)+\lambda \cdot \one  )$ $= (u_1, u_1+\lambda \cdot \one, u_2',u_2'+\lambda \cdot \one, u_3',u_3'+\lambda \cdot \one)$,
where $u_2'=\Phi_2(u_2)$ and $u_3'=\Phi_3(u_3)$. Thus, 
\begin{equation*}
\begin{split}
A_0&=\{ (u_1, u_1, u_2',u_2', u_3',u_3'):(u_1, u_2', u_3') \in H^{1,0,t_3}\},\\
A_1&=\{ (u_1, u_1+ \one, u_2',u_2' +\one, u_3',u_3'+ \one)):(u_1, u_2', u_3') \in H^{1,0,t_3}\}.\\
\end{split}
\end{equation*}
Since $H^{1,0,t_3}$ is linear, it is clear that if we take any two codewords from $H$, then their addition belongs to one of the blocks $A_0$ and $A_1$. Therefore, $H$ is linear.
\end{proof}

\begin{theorem}\label{linearity:Z2Z4Z8}
The codes $H^{1,0,t_3}$ of type $(\alpha_1,\alpha_2,\alpha_3;1,0,t_3)$, with $t_3 \geq 1$, are the only $\Z_2\Z_4\Z_8$-linear Hadamard codes with $\alpha_1\not =0$, $\alpha_2\not =0$, and $\alpha_3\not =0$, which are linear.
\end{theorem}
\begin{proof}
By Proposition \ref{Th:linear}, we have that the codes $H^{1,0,t_3}$, with $t_3 \geq 1$, are linear.

First, we show that $H^{2,0,1}=\Phi({\cH}^{2,0,1})$ is nonlinear. By Example \ref{ex:MatricesA}, the code  ${\cH}^{2,0,1}$ is generated by the matrix $A^{2,0,1}$ given in (\ref{eq:A201}). 
% $$
% \footnotesize
% \left(\begin{array}{cccc|cccccc|cccccccccccc}
% 1 &1 &1 &1 &2 &2 &2 &2 &2 &2 &4 &4 &4 &4 &4 &4 &4 &4 &4 &4 &4 &4 \\
% 0  &1 &0  &1 &0 &2 &1 &1 &1 &1 &0 &2 &4 &6 &1 &1 &1 &1 &1 &1 &1 &1 \\
% 0 &0 &1 &1 &1 &1 &0 &1 &2 &3 &1 &1 &1 &1 &0 &1 &2 &3 &4 &5 &6 &7\\ 
% \end{array}\right).
% $$
Let $\uu=(u_1\mid u_2\mid u_3)$ and $\vv=(v_1 \mid v_2\mid v_3)$ be the second and the third row vectors of $A^{2,0,1}$.
Now, by Corollary \ref{coro1},
\begin{align*}
    &\Phi(\uu)+\Phi(\vv)\\
    &=(u_1+v_1, \Phi_2(u_2)+\Phi_2(v_2), \Phi_3(u_3)+\Phi_3(v_3))\\
    &=(u_1+v_1 , \Phi_2(u_2+v_2-2(u_2 \odot v_2)), \Phi_3(u_3+v_3-2(u_3 \odot v_3)))\\
    &=\Phi(u_1+v_1\mid  u_2+v_2-2(u_2 \odot v_2)\mid u_3+v_3-2(u_3 \odot v_3))\\
    &=\Phi((u_1+v_1 \mid u_2+v_2\mid  u_3+v_3)-(0,0,0,0\mid  2(u_2 \odot v_2)\mid  2(u_3 \odot v_3))).
\end{align*}
Therefore, since $(u_1+v_1 \mid u_2+v_2\mid  u_3+v_3)=\uu + \vv \in  {\cH}^{2,0,1}$, we just need to show that $$\ww=(0,0,0,0\mid  2(u_2 \odot v_2)\mid 2(u_3 \odot v_3)) \notin {\cH}^{2,0,1}.$$ By Corollary \ref{coro2},  $\Phi_2(2(u_2 \odot v_2))=(\zero, \zero,\zero, \one, \zero, \one)$, where the vectors $\zero$ and $\one$ are of length $2$, so $\wt_H(\Phi_2(2(u_2 \odot v_2)))=4$. By Corollary \ref{coroNew1}, $\Phi_3(2(u_3 \odot v_3))=(\zero, \zero,\zero, \zero, \zero, \phi_3(2), \zero, \phi_3(2), \zero, \phi_3(2), \zero, \phi_3(2))$, where $\zero=(0,0,0,0)$, so $\wt_H(\Phi_3(2(u_3 \odot v_3)))=8$. Therefore, $\wt_H(\Phi(\ww))=\wt_H(\Phi_2(2(u_2 \odot v_2)))+\wt_H(\Phi_3(2(u_3 \odot v_3)))=4+8=12<32$, and hence $\Phi(\ww) \notin H^{2,0,1}$ since the minimum distance of $H^{2,0,1}$ is $32$.

Second, we prove that if $H^{\ell,0,1}$, where $\ell \geq 2$, is nonlinear, then $H^{\ell+1,0,1}$ is also nonlinear. Assume that $H^{\ell+1,0,1}$ is linear. Then, by construction (\ref{eq:recGenMatrix1}), for any $\uu=(u_1\mid u_2\mid u_3)$, $\vv=(v_1\mid v_2\mid v_3) \in \mathcal{H}^{\ell,0,1}$, we have that 
\begin{equation*}
  \begin{array}{c}
 \Bar{\uu}=(u_1, u_1\mid x_1, u_2,\stackrel{4}{\dots}, u_2 \mid x_2, u_3,\stackrel{8}{\dots}, u_3)\in \mathcal{H}^{\ell+1,0,1}, \\
 \Bar{\vv}=(v_1, v_1\mid x'_1, v_2,\stackrel{4}{\dots}, v_2 \mid x'_2, v_3,\stackrel{8}{\dots}, v_3)\in \mathcal{H}^{\ell+1,0,1},\\
\end{array}  
\end{equation*}
where $x_{i-1}$, $x'_{i-1} \in (2\Z_{2^i})^{\ell}$ for $i \in \{2,3\}$.
Since $H^{\ell+1,0,1}$ is linear, $\Phi(\Bar{\uu})+\Phi(\Bar{\vv})\in H^{\ell+1,0,1}$. Again, by construction (\ref{eq:recGenMatrix1}), we have that $\Phi(\Bar{\uu})+\Phi(\Bar{\vv})=\Phi((a_1,a_1\mid y_1, a_2,\stackrel{4}{\dots},a_2\mid y_2, a_3,\stackrel{8}{\dots},a_3 )+\lambda(\zero,\one \mid \one, \zero, \one, \mathbf{2}, \mathbf{3} \mid \one, \zero, \one, \dots, \mathbf{7}))\in H^{\ell+1,0,1}$, for some $\mathbf{a}=(a_1\mid a_2\mid a_3) \in {\cH}^{\ell,0,1}$, $y_{i-1} \in (2\Z_{2^i})^{\ell}$ for $i \in \{2,3\}$, and $\lambda \in \Z_8$. Considering the coordinates in positions $1$, $4$, and $9$ of $\Bar{\uu}$ and $\Bar{\vv}$, we have that $\Phi(\uu)+\Phi(\vv)=\Phi(\mathbf{a}) \in H^{\ell, 0, 1}$,
and then $H^{\ell,0,1}$ is linear, which is a contradiction.

Third, we show that that $H^{1,1,1}=\Phi({\cH}^{1,1,1})$ is nonlinear. We consider the matrix $A^{1,1,1}$ given in (\ref{eq:A111}) and, by using the same argument as before, we can see that $\Phi(\uu)+\Phi(\vv)\not \in H^{1,1,1}$, where $\uu$ and $\vv$ are the second and the third row vectors of $A^{1,1,1}$. 
 
 Finally, if  $H^{t_1, \ell,1}$, with $\ell \geq 0$, (respectively, $H^{t_1, t_2, \ell}$, with $\ell \geq 1$) is nonlinear, then as above we can show that $H^{t_1, \ell+1,1}$ (respectively, $H^{t_1, t_2, \ell+1}$) is also nonlinear, and hence the result follows.
\end{proof}

\section{Kernel of $\Z_2\Z_4\Z_8$-linear Hadamard codes}\label{Sec:Kernel}

The computation of the kernel and its dimension for $\Z_4$-linear Hadamard codes is given in \cite{Kro:2001:Z4_Had_Perf,PRV06}. In general, for $\Z_{p^s}$-linear Hadamard codes with $s\geq 2$ and $p$ prime, the results on the kernel are given in \cite{KernelZ2s} when $p=2$ and in \cite{HadamardZps} when $p\geq 3$. Regarding codes over mixed alphabets, it is proved that the kernel of a nonlinear $\Z_p\Z_{p^2}$-linear Hadamard code coincides with the Gray map image of the elements of order at most $p$ of the corresponding $\Z_p\Z_{p^2}$-additive code of type $(\alpha_1,\alpha_2;t_1,t_2)$ with $\alpha_1\not =0, \alpha_2\not =0$ in \cite{PRV06} and \cite{ZpZp2Classification}, for $p=2$ and $p\geq 3$ prime, respectively. Thus, the dimension of the kernel for these codes is equal to $t_1+t_2$. In this section, we generalize some of these results and show that the dimension of the kernel for nonlinear $\Z_2\Z_4\Z_8$-linear Hadamard codes $H^{t_1,t_2,t_3}$ %\cite{Z2Z4Z8Construction} 
of type $(\alpha_1,\alpha_2,\alpha_3;t_1,t_2,t_3)$ with $\alpha_1\not =0$, $\alpha_2\not =0$, and $\alpha_3\not =0$  is equal to $t_1+t_2+t_3$. 

\medskip
When we include all the elements of  $\Z_{2^i}$, where $1\leq i\leq 3$,  as coordinates of a vector, we place them in increasing order. 
For a set $S \subseteq \Z_{2^i}$ and $\lambda \in \Z_{2^i}$, where $i\in \{1,2,3\}$, we define $\lambda S=\{\lambda j: j\in S\}$ and $S+\lambda=\{j+\lambda: j\in S\}$. As before, when including all the elements in those sets as coordinates of a vector, we place them in increasing order. 
For example, 
$2\Z_8=\{0,4,6,8\}$, $(\Z_4,\Z_4)=(0,1,2,3,0,1,2,3) \in \Z_4^8$ and $(\Z_2 \mid \Z_4\mid 2\Z_8, 4\Z_8)=(0,1 \mid 0,1,2,3\mid 0,2,4,6, 0,4) \in \Z_2^2 \times \Z_4^4 \times \Z_8^6$.

\begin{theorem}
Let $t_1\geq 1$, $t_2\geq 0$, and $t_3\geq 1$ be integers.  Let ${\cH}= {\cH}^{t_1,t_2,t_3}$ be the $\Z_2\Z_4\Z_8$-additive Hadamard code of type $(\alpha_1, \alpha_2,\alpha_3; t_1,t_2,t_3)$ such that $\Phi({\cH})$ is nonlinear.  Let ${\cH}_2$ be the subcode of ${\cH}$ which contains all the codewords of order at most two. Then, $K(\Phi({\cH}))=\Phi({\cH}_2)$.
\end{theorem}
\begin{proof}
By Corollary \ref{lemma2}, for all 
$\mathbf{b}=(b_1\mid b_2\mid b_3)\in\mathcal{H}_2$ and
$\uu=(u_1\mid u_2\mid u_3)\in\cH$, we have that
$\Phi(\mathbf{b})+\Phi(\mathbf{u})=(b_1+u_1, \Phi_2(b_2)+\Phi_2(u_2), \Phi_3(b_3)+\Phi_3(u_3))=(b_1+u_1, \Phi_2(b_2+u_2), \Phi_3(b_3+u_3))=\Phi(b_1+u_1\mid b_2+u_2 \mid b_3+u_3)\in \Phi(\cH)$, so $\Phi({\cH}_2)\subseteq K(\Phi(\cH))$. 

Let $\Phi(\bar \uu)\in K(\Phi(\cH))$, where $\bar \uu \not =\zero$. 
 We prove that $o(\bar \uu)=2$, and thus $K(\Phi({\cH})) \subseteq \Phi({\cH}_2)$. Let $\{\mathbf{b}_1,\dots, \mathbf{b}_{t_1}\}$ (respectively, $\{\mathbf{c}_1,\dots, \mathbf{c}_{t_2}\}$) be the collection of rows of order $8$ (respectively, $4$) of the generator matrix $A^{t_1,t_2, t_3}$ of the code ${\cH}$.
 Then, we can write $\bar \uu=\uu+\uu_0$, where $\uu=\sum_{i=1}^{t_1}\lambda_i \mathbf{b}_i+\sum_{j=1}^{t_2}\mu_j \mathbf{c}_j$, $\lambda_i \in \{0,1,2,3\}\subseteq \Z_8$, $\mu_j \in \{0,1\} \subseteq \Z_8$, and $o(\uu_0)\leq 2$. By Corollary \ref{lemma2}, $\Phi(\bar \uu)=\Phi(\uu)+\Phi(\uu_0)$.  Since $\Phi(\bar \uu)\in K(\Phi(\cH))$ and $\Phi({\cH}_2)\subseteq K(\Phi(\cH))$, we have that $\Phi(\uu)\in K(\Phi(\cH))$.  Assume that  $o(\bar \uu)>2$, so $o(\uu)>2$. 
 In order to obtain a contradiction, we just need to find an element $\tilde{\uu}\in \mathcal{H}$ such that $\Phi(\uu)+\Phi(\tilde{\uu})\not\in \Phi(\cH)$. 
 
 Let $\ww_2=(w_1\mid w_2\mid w_3)$ be the second row of  $A^{t_1,t_2,t_3}$. Note that the number of ones in $w_3$ is $m=8^{t_1-1}4^{t_2}2^{t_3-1}$. Then, by Proposition \ref{relation:t1a1a3}, we have that $m=n/8$, where $n=\alpha_1+2\alpha_2+ 4\alpha_3$ is the length of $\Phi(\cH)$. Let $I$ be the set of coordinate positions corresponding to these ones in $w_3$. 

Assume that $o(\uu)=8$ and $\uu \notin \{\ww_2, 3\ww_2\}$. Let $\vv=2\ww_2=(v_1\mid v_2 \mid v_3)$. Note that $v_3$ contains the element $2$ in the coordinates corresponding to $I$, and $0$ or $4$ in the other coordinates. By construction, in the coordinates corresponding to $I$, either $u_3$ is a permutation of the vector $(\Z_8, \stackrel{m/8}{\dots}, \Z_8)$ or it is a permutation of $(\Z_8{\setminus}2\Z_8, \stackrel{m/4}{\dots}, \Z_8{\setminus}2\Z_8)$.  Now, by Corollary \ref{coro1},
\begin{align*}
    &\Phi(\uu)+\Phi(\vv)\\
    &=(u_1+v_1, \Phi_2(u_2)+\Phi_2(v_2), \Phi_3(u_3)+\Phi_3(v_3))\\
    &=(u_1+v_1 , \Phi_2(u_2+v_2), \Phi_3(u_3+v_3-2(u_3 \odot v_3)))\\
    &=\Phi(u_1+v_1\mid  u_2+v_2\mid u_3+v_3-2(u_3 \odot v_3))\\
    &=\Phi((u_1+v_1 \mid u_2+v_2\mid  u_3+v_3)-(\zero \mid  \zero \mid  2(u_3 \odot v_3)))\\
    &=\Phi(\uu+\vv-\mathbf{x}),
\end{align*}   
where $\mathbf{x}=(\zero \mid  \zero \mid  2(u_3 \odot v_3))$. By Corollary  \ref{coroNew1}, $\wt_H(\Phi(\mathbf{x}))=\wt_H(\Phi_3(2(u_3 \odot v_3)))=\frac{m}{8}\cdot 16
%(or~ \frac{m}{4}\cdot 8 )
=2m$ if $u_3$ is a permutation of $(\Z_8, \stackrel{m/8}{\dots}, \Z_8)$, and $\wt_H(\Phi(\mathbf{x}))=\frac{m}{4}\cdot 8
=2m$ otherwise. Since $\wt_H(\Phi(\mathbf{x}))=2m=n/4$ and the minimum weight of $\Phi({\cH})$ is $n/2$, we
have that $\Phi(\mathbf{x}) \notin \Phi({\cH})$. %by corollary \ref{relation:t1a1a3}. 
Therefore, $\Phi(\uu) \notin K(\Phi({\cH}))$, which is a contradiction.

Assume that $o(\uu)=8$ and $\uu \in \{\ww_2, 3\ww_2\}$. Let $\ww_3=(w'_1\mid w'_2\mid w'_3)$ be the third row of  $A^{t_1,t_2,t_3}$.  First, consider that $t_1>1$. Let $k$ be the number of ones in $w'_3$ outside the coordinate positions of $I$. Then, $k=4\cdot8^{t_1-2}4^{t_2}2^{t_3-1}=m/2.$ Let $\mathbf{z}=2\ww_3=(z_1\mid z_2\mid z_3)$. Note that in the coordinates corresponding to $I$, $z_3$ contains every element of $2\Z_8$ exactly $m/4$ times. Now, $\Phi(\uu)+\Phi(\mathbf{z})=\Phi(\uu+\mathbf{z}-\mathbf{x}_1)$, where $\mathbf{x}_1=(\zero \mid  \zero \mid  2(u_3 \odot z_3))$. 
By Corollary  \ref{coroNew1}, 
$$
\wt_H(\Phi(\mathbf{x}_1))=\wt_H(\Phi_3(2(u_3 \odot z_3)))= \left\{
 \begin{tabular}{l l l}
 $\frac{k}{4}\cdot 8=m$ & if $\uu=\ww_2$,\\
 $\frac{k}{4}\cdot 8$+ $\frac{m}{4}\cdot 8=3m$ & if $\uu=3\ww_2.$
\end{tabular}
  \right.
$$
Since $m=n/8$ and the minimum weight of $\Phi({\cH})$ is $n/2$, we have that $\Phi(\mathbf{x}_1) \notin \Phi({\cH})$. Therefore, $\Phi(\uu) \notin K(\Phi({\cH}))$, which is a contradiction. %Let $o(\uu)=8$ such that $\uu \in \{5\ww, 7\ww\}$. Since $\Phi(5\ww)=\Phi(\ww)+ \Phi(4\ww)$, $\Phi(7\ww)=\Phi(3\ww)+ \Phi(4\ww)$, and $4\ww$ is a vector of order 2, we have that $\Phi(\uu)+ \Phi(\mathbf{z})= \Phi(\uu+\mathbf{z}-(\zero \mid  \zero \mid  2(u_3 \odot z_3))$. So, as above, we get a contradiction.
Second, consider that $t_1=1$. Then, $m=\alpha_3$. 
Note that the number of ones in $w_2$ is $m_2=4^{t_2}2^{t_3-1}$. Then, by Proposition \ref{relation:t1a1a3}, we have that $m_2=n/8$. Let $I_2$ be the set of coordinate positions to these ones in $w_2$. Let $\mathbf{y}=\ww_2+\ww_3=(y_1\mid y_2\mid y_3)$.
Then, we have that $y_3=(\Z_8{\setminus 2\Z_8},\stackrel{m/4}{\dots}, \Z_8{\setminus 2\Z_8})$, and in the coordinates corresponding to $I_2$, $y_2$ is a permutations of the vector $(\Z_4, \stackrel{m_2/4}{\dots}, \Z_4)$.
Now, $\Phi(\uu)+\Phi(\mathbf{y})=\Phi(\uu+\mathbf{y}-\mathbf{x}_2)$, where $\mathbf{x}_2=(\zero \mid  2(u_2 \odot y_2) \mid  2(u_3 \odot y_3))$. 
By Corollary  \ref{coroNew1},
\begin{align*}
    \wt_H(\Phi(\mathbf{x}_2)) & = \wt_H(\Phi_2(2(u_2 \odot y_2)))+\wt_H(\Phi_3(2(u_3 \odot y_3)))=\\
    & = \frac{m_2}{4}\cdot  2\cdot 2+\frac{m}{4}\cdot 8 \mbox{ if } \uu \in \{\ww_2, 3\ww_2\}.
\end{align*}
Since $m_2=m=n/8$ and the minimum weight of $\Phi({\cH})$ is $n/2$, we have that $\Phi(\mathbf{x}_2) 
\notin \Phi({\cH})$. Therefore, $\Phi(\uu) \notin K(\Phi({\cH}))$, which is a contradiction.

   Assume that $o(\uu)=4$. Recall that $\vv=2\mathbf{w}_2$, $\zz = 2\mathbf{w}_3$, and $\mathbf{y}=\mathbf{w}_2 + \mathbf{w}_3$. First, consider that $t_1>1$. By construction, in the coordinates corresponding to $I$, either $u_3$  contains every element of $2\Z_8$ exactly $m/4$ times or $u_3$ is equal to $\two$. In the first case, we consider 
  $\Phi(\uu)+\Phi(\vv)=\Phi(\uu+\vv-\mathbf{x}_3)$, where $\mathbf{x}_3=(\zero \mid  \zero \mid  2(u_3 \odot v_3))$.  In the second case, we consider $\Phi(\uu)+\Phi(\mathbf{z})=\Phi(\uu+\mathbf{z}-\mathbf{x}_4)$, where $\mathbf{x}_4=(\zero \mid  \zero \mid  2(u_3 \odot {z}_3)$. Then, in both cases, $\wt_H(\Phi(\mathbf{x}_3))=\wt_H(\Phi(\mathbf{x}_4))=\frac{m}{4}\cdot 8=2m=n/4$. 
  %Since the minimum weight of $\Phi({\cH})$ is $n/2$, we have that $\Phi(\mathbf{x}_3), \Phi(\bar{\mathbf{x}}_3)  \notin \Phi({\cH})$, which is a contradiction. 
  Second, consider that $t_1=1$. Note that $m=\alpha_3$, so $u_3$  is a permutation of $(2\Z_8,\stackrel{m/4}{\dots}, 2\Z_8)$ or $u_3=\two$. In the first case, we consider again $\Phi(\uu)+\Phi(\vv)=\Phi(\uu+\vv-\mathbf{x}_3)$, and in the second case,  
  $\Phi(\uu)+\Phi(\mathbf{y})=\Phi(\uu+\mathbf{y}-\mathbf{x}_5)$, where $\mathbf{x}_5=(\zero \mid  \zero \mid  2(u_3 \odot {y}_3)$. In both cases, $\wt_H(\Phi(\mathbf{x}_3))= \wt_H(\Phi(\mathbf{x}_5))=\frac{m}{4}\cdot 8=2m=n/4$. For all these cases, since the minimum weight of $\Phi({\cH})$ is $n/2$, we have that $\Phi(\mathbf{x}_3)$, $\Phi(\mathbf{x}_4)$, $\Phi(\mathbf{x}_5)  \notin \Phi({\cH})$, which is a contradiction.
  This completes the proof. 
\end{proof}

\begin{corollary}\label{kernel:Z2Z4Z8}
Let $t_1\geq 1$, $t_2\geq 0$, and $t_3\geq 1$ be integers.	Let ${\cH}= {\cH}^{t_1,t_2,t_3}$ be the $\Z_2\Z_4\Z_8$-additive Hadamard code of type $(\alpha_1, \alpha_2,\alpha_3; t_1,t_2,t_3)$ such that $\Phi({\cH})$ is nonlinear.  Let $\mathbf{w}_k$ be the $k$th row of $A^{t_1,t_2,t_3}$ and $Q=\lbrace (o(\mathbf{w}_k)/2)\mathbf{w}_k\rbrace_{k=1}^{t_1+t_2+t_3}$. Then,
	$\Phi(Q)$ is a basis of $K(\Phi({\cH}))$ and $\kernel(\Phi({\cH}))=t_1+t_2+t_3$.
\end{corollary}

\section{Classification results}\label{Sec:classification}

The classification of $\Z_2\Z_4$-linear Hadamard codes of type $(\alpha_1,\alpha_2;t_2,t_3)$, (or equivalently,  $\Z_2\Z_4\Z_8$-linear Hadamard codes of type $(\alpha_1, \alpha_2,0;0,t_2,t_3)$) with $\alpha_1\neq 0$ and $\alpha_2\neq 0$, of length $2^t$, for any $t\geq 3$, using the rank or the dimension of the kernel is shown in \cite{PRV06}.
 For $\Z_4$-linear Hadamard codes of type $(n; t_2,t_3)$ (or equivalently,  $\Z_2\Z_4\Z_8$-linear Hadamard codes of type $(0, n,0;0,t_2,t_3)$), the classification is shown in \cite{Kro:2001:Z4_Had_Perf,PRV06}. In \cite{KV2015}, it is shown that each $\Z_4$-linear Hadamard code is  equivalent to a  $\Z_2\Z_4$-linear Hadamard code with $\alpha_1\neq 0$ and $\alpha_2\neq 0$. %{\color{red}\xout{, so indeed there are only $\lfloor t/2\rfloor$ nonequivalent $\Z_2\Z_4$-linear Hadamard codes of length $2^t$}}. 
 Some partial results on the classification of $\Z_{2^s}$-linear Hadamard codes of length $2^t$, for any $t\geq 3$ and $s>2$, are proved in \cite{KernelZ2s,EquivZ2s}; and in general for $\Z_{p^s}$-linear Hadamard codes of length $p^t$, for any $t\geq 2$, $s\geq 2$, and $p\geq 3$  prime, in \cite{HadamardZps,ZpsEquivalance}. For any $t\geq 2$, the full classification of $\Z_p\Z_{p^2}$-linear Hadamard codes of length $p^t$, with $\alpha_1\neq 0$, $\alpha_2\neq 0$, and $p\geq 3$ prime, is given in \cite{ZpZp2Construction,ZpZp2Classification}, by using just the dimension of the kernel. 
 % In this section, we give some partial classification results by comparing the $\Z_2\Z_4\Z_8$-linear Hadamard codes $H^{t_1,t_2,t_3}$ %\cite{Z2Z4Z8Construction} 
 % of length $2^t$ with $\Z_2\Z_4$-linear and $\Z_{2^s}$-linear Hadamard codes \cite{KernelZ2s}, with $s\geq 2$, of length $2^t$.
  In this section, we present several families of infinite such nonlinear $\Z_2\Z_4\Z_8$-linear Hadamard codes $H^{t_1,t_2,t_3}$, which are not equivalent to any other constructed $\Z_2\Z_4\Z_8$-linear Hadamard code, nor to any $\Z_2\Z_4$-linear Hadamard code, nor to any previously constructed $\Z_{2^s}$-linear Hadamard code \cite{KernelZ2s}, with $s\geq 2$, with the same length $2^t$. Moreover, for $3 \leq t \leq 11$, we give a complete classification by providing the exact amount of nonequivalent such codes.
  
Next, we recall the construction of $\Z_{2^s}$-linear Hadamard codes with $s\geq 2$, given in \cite{KernelZ2s}. From \cite{KernelZ2s,EquivZ2s},  we also recall for which types these codes are linear, what are their kernel dimensions when they are nonlinear, and which of them are equivalent to each other. Then,  we recall the results on the rank of $\Z_8$-linear and $\Z_2\Z_4$-linear Hadamard codes given in \cite{fernandez2019mathbb} and \cite{PRV06}, respectively.  Finally, by using these previous results, we present some classification results for $\Z_2\Z_4\Z_8$-linear Hadamard codes $H^{t_1,t_2,t_3}$ %\cite{Z2Z4Z8Construction} 
of length $2^t$, by comparing them with any other $\Z_2\Z_4\Z_8$-linear Hadamard codes, and with  any  $\Z_2\Z_4$-linear and $\Z_{2^s}$-linear Hadamard codes \cite{KernelZ2s}, where $s\geq 2$, of length $2^t$. Moreover, for $3 \leq t \leq 11$, we present a complete classification of $\Z_2\Z_4\Z_8$-linear Hadamard codes $H^{t_1,t_2,t_3}$ of length $2^t$ by providing the exact amount of nonequivalent such codes.

First, we recall the construction of  $\Z_{2^s}$-linear Hadamard codes with $s\geq 2$ and some results, given in \cite{KernelZ2s,EquivZ2s,fernandez2019mathbb}. Let $T_i=\lbrace j\cdot 2^{i-1}\, :\, j\in\lbrace0,1,\dots,2^{s-i+1}-1\rbrace \rbrace$ for all $i \in \{1,\ldots,s \}$.
Note that $T_1=\lbrace0,\dots,2^{s}-1\rbrace$. Let $t_1,\dots,t_s$ be non-negative integers with $t_1\geq1$. Consider the matrix $\bar{A}^{t_1,\dots,t_s}$ whose columns are exactly all the vectors of the form $\mathbf{z}^T$, $\mathbf{z}\in\lbrace1\rbrace\times T_1^{t_1-1}\times T_{2}^{t_2}\times\cdots\times T_s^{t_s}$.

% \begin{example}
% Some matrices for $s=3$ are the following:
% $$
% \arraycolsep=1.4pt\def\arraystretch{}
% \bar{A}^{1,0,1}=\left(\begin{array}{cc}
% 1 & 1\\
% 0 & 4\\
% \end{array}\right),\quad \bar{A}^{1,1,0}=\left(\begin{array}{cccc}
% 1 & 1 & 1 & 1\\
% 0 & 2 & 4 & 6\\
% \end{array}\right),\quad \bar{A}^{2,0,0}=\left(\begin{array}{cccccccc}
% 1 & 1 & 1 & 1 & 1 & 1 & 1 & 1\\
% 0 & 1 & 2 & 3 & 4 & 5 & 6 & 7\\
% \end{array}\right),
% $$
% $$
% \arraycolsep=1.4pt\def\arraystretch{}
% \bar{A}^{1,1,1}=\left(\begin{array}{ccccccccc}
% 1 & 1 & 1 & 1 & & 1 & 1 & 1 & 1\\
% 0 & 2 & 4 & 6 & & 0 & 2 & 4 & 6\\
% 0 & 0 & 0 & 0 & & 4 & 4 & 4 & 4\\
% \end{array}\right),\quad \bar{A}^{2,0,1}=\left(\begin{array}{ccccccccccccccccc}
% 1 & 1 & 1 & 1 & 1 & 1 & 1 & 1 & & 1 & 1 & 1 & 1 & 1 & 1 & 1 & 1 \\
% 0 & 1 & 2 & 3 & 4 & 5 & 6 & 7 & & 0 & 1 & 2 & 3 & 4 & 5 & 6 & 7 \\
% 0 & 0 & 0 & 0 & 0 & 0 & 0 & 0 & & 4 & 4 & 4 & 4 & 4 & 4 & 4 & 4\\
% \end{array}\right),
% $$
% $$
% \arraycolsep=1.4pt\def\arraystretch{}
% \bar{A}^{2,1,0}=\left(\begin{array}{ccccccccccccccccccccccccccccccccccc}
% 1 & 1 & 1 & 1 & 1 & 1 & 1 & 1 & & 1 & 1 & 1 & 1 & 1 & 1 & 1 & 1 & & 1 & 1 & 1 & 1 & 1 & 1 & 1 & 1 & & 1 & 1 & 1 & 1 & 1 & 1 & 1 & 1\\
% 0 & 1 & 2 & 3 & 4 & 5 & 6 & 7 & & 0 & 1 & 2 & 3 & 4 & 5 & 6 & 7 & & 0 & 1 & 2 & 3 & 4 & 5 & 6 & 7 & & 0 & 1 & 2 & 3 & 4 & 5 & 6 & 7\\
% 0 & 0 & 0 & 0 & 0 & 0 & 0 & 0 & & 2 & 2 & 2 & 2 & 2 & 2 & 2 & 2 & & 4 & 4 & 4 & 4 & 4 & 4 & 4 & 4 & & 6 & 6 & 6 & 6 & 6 & 6 & 6 & 6\\
% \end{array}\right).
% $$
% \end{example}

Any matrix $\bar{A}^{t_1,\dots,t_s}$ can also be obtained by applying the
following recursive construction. We start with $\bar{A}^{1,0,\dots,0}=(1)$. Then, if
we have a matrix $\bar{A}=\bar{A}^{t_1,\dots,t_s}$, for any $i\in \{1,\ldots,s\}$, we may construct the matrix
\begin{equation}\label{eq:recGenMatrix}
\bar{A}_i=
\left(\begin{array}{cccc}
\bar{A} & \bar{A} &\cdots & \bar{A} \\
0\cdot \mathbf{2^{i-1}}  & 1\cdot \mathbf{2^{i-1}} & \cdots & (2^{s-i+1}-1)\cdot \mathbf{2^{i-1}}  \\
\end{array}\right).
\end{equation}
Finally, by permuting the rows of $\bar{A}_i$, we obtain a matrix $\bar{A}^{t'_1,\ldots,t'_s}$, where $t'_j=t_j$ for $j\not=i$ and $t'_i=t_i+1$. Note that any permutation of columns of $\bar{A}_i$ gives also a matrix $\bar{A}^{t_1',\dots,t_s'}$.

We consider that the matrices $\bar{A}^{t_1,\ldots,t_s}$ are constructed recursively starting from $\bar{A}^{1,0,\ldots,0}$ in the following way. First, we add $t_1-1$ rows of order $2^s$, up to obtain $\bar{A}^{t_1,0,\ldots,0}$; then $t_2$ rows of order $2^{s-1}$ up to generate $\bar{A}^{t_1,t_2,\ldots,0}$; and so on, until we add $t_s$ rows of order $p$ to achieve $\bar{A}^{t_1,\ldots,t_s}$. See \cite{KernelZ2s} for examples.
Let $\bar{\mathcal{H}}^{t_1,\dots,t_s}$ be the $\Z_{2^s}$-additive code of type $(n;t_1,\dots,t_s)$  generated by $\bar{A}^{t_1,\dots,t_s}$, where $t_1,\dots,t_s$ are non-negative integers with $t_1\geq1$. 
Let $\bar{H}^{t_1,\dots,t_s}=\Phi(\bar{\mathcal{H}}^{t_1,\dots,t_s})$ be the corresponding $\Z_{2^s}$-linear Hadamard code.

Let $\mathbf{w}_i$ be the $i$th row of $\bar{A}^{t_1,\ldots,t_s}$, $1\leq i\leq t_1+\cdots+t_s$.
By construction, $\mathbf{w}_1=\one$ and  $o(\mathbf{w}_i)\leq o(\mathbf{w}_j)$ if $i>j$. We define $\sigma\in\lbrace1,\dots,s\rbrace$ as the integer such that $o(\textbf{w}_2)=2^{s+1-\sigma}$. For $\bar{\mathcal{H}}^{1,0,\dots,0}$, we define $\sigma=s$. Note that $\sigma=1$ if $t_1>1$, and $\sigma=\min \lbrace i : t_i>0, i\in\lbrace2,\dots,s\rbrace\rbrace$ if $t_1=1$.

\begin{theorem}\cite{KernelZ2s}\label{Th:Z2Had}
Let $t_1,\dots,t_s$ be non-negative integers with with $s\geq 2$ and $t_1\geq1$.
The $\Z_{2^s}$-linear code $\bar{H}^{t_1,\dots,t_s}$ of type $(n;t_1,\dots,t_s)$
is a binary Hadamard code of length $2^t$, with
$t=\left(\sum_{i=1}^{s}(s-i+1)\cdot t_i\right)-1$ and $n=2^{t-s+1}$.
\end{theorem}

% \begin{theorem}\cite{KernelZ2s}\label{theorem:linealZ2s}
% The codes $\bar{H}^{1,0,\dots,0,t_s}$ and $\bar{H}^{1,0,\dots,0,1,t_s}$, with $s\geq 3$ and $t_s\geq0$, are the only $\Z_{2^s}$-linear Hadamard codes which are linear.
% \end{theorem}

\begin{theorem}\cite{KernelZ2s}\label{Teo:kernel1} 
 The  $\Z_{2^s}$-linear Hadamard code $\bar{H}^{t_1,\dots,t_s}$ of type $(n; t_1,$ $\dots, t_s)$ is linear if and only if $(t_1,\dots,t_s)\in\{(1,0,\dots,0,t_s),$ $(1,0,\dots,0,1,t_s)\}$. Moreover, if $\bar{H}^{t_1,\dots,t_s}$ is nonlinear, then $\kernel(\bar{H}^{t_1,\dots,t_s})=\sigma+\sum_{i=1}^{s}t_i$.
\end{theorem}

\begin{theorem}\cite{fernandez2019mathbb}\label{rankZ2s}
   Let $\bar{H}^{t_1,\dots,t_s}$ be the $\Z_{2^s}$-linear Hadamard code of type $(n; t_1,$ $\dots, t_s)$. Then, $\rank(\bar{H}^{t_1,\dots,t_s+1})=1+\rank(\bar{H}^{t_1,\dots,t_s})$.
\end{theorem}

\begin{theorem}\cite{fernandez2019mathbb}\label{rankZ8}
Let $\bar{H}^{t_1,t_2,t_3}$ be the $\Z_8$-linear Hadamard code of type $(n;t_1,t_2,t_3)$. Then,
$$
\rank(\bar{H}^{t_1,t_2,t_3})=\frac{t^4_1}{24}-\frac{t^3_1}{12}+\frac{35t^2_1}{24}+\frac{7t_1}{12}+\frac{t_2}{2}(t^2_1+t_1+t_2+1)+t_3+1.
$$
\end{theorem}

\begin{theorem}\cite{EquivZ2s}\label{theo:equi}
Let $\bar{H}^{t_1,\dots,t_s}$ be the $\Z_{2^s}$-linear Hadamard code, with $s\geq 2$ and $t_s\geq 1$.
Then,  for all $\ell\in\{1,\dots,t_s\}$, $\bar{H}^{t_1,\dots,t_s}$  is permutation equivalent to the
$\Z_{2^{s+\ell}}$-linear Hadamard code
$\bar{H}^{1,\zero^{\ell-1},t_1-1,t_2,\dots,t_{s-1},t_s-\ell}.$
\end{theorem}

 %\medskip
Let $t'_1,\dots,t'_{s'}$ be non-negative integers with $t'_1\geq 2$ if $s'\geq 3$, or $t'_1\geq 1$ if $s'=2$. 
Let $C_H(t'_1,\dots,t'_{s'})=[H_1=\bar{H}^{t'_1,\dots,t'_{s'}},H_2,\dots, H_\rho]$ be the sequence of all $\Z_{2^{s}}$-linear Hadamard codes $\bar{H}^{t_1,\dots,t_s}$ of length $2^t$, where $t=$ $(\sum_{i=1}^{s}(s-i+1)\cdot t_i)-1$, that are permutation equivalent to $\bar{H}^{t'_1,\dots,t'_{s'}}$ by Theorem \ref{theo:equi}.
% We denote by $C_H(t_1,\dots,t_s)[i]$ the $i$th code $H_i$ in the sequence, for $1\leq i\leq t_s+1$. We consider that the order of the codes in $C_p(t_1,\dots,t_s)$ is the following:
% \begin{equation}\label{eq:defCH}
% 	C_p(t_1,\dots,t_s)[i]=
% 		\begin{cases}
% 	H_p^{t_1,\dots,t_{s}} &\text{if } i= 1,\\
% 	H_p^{1,\zero^{i-2},t_1-1,t_2,\dots,t_{s-1},t_s-i+1} &\text{otherwise.}
% \end{cases}
% \end{equation}
 We refer to $C_H(t'_1,\dots,t'_s)$ as the chain of equivalences of $\bar{H}^{t'_1,\dots,t'_s}$. If we focus on nonlinear $\Z_{2^s}$-linear Hadamard codes with $s\geq 3$, then we can assume that $t'_1\geq 2$ \cite{ZpsEquivalance}.

\begin{proposition}\cite{EquivZ2s}\label{Z2sEqui}
Let $t_1, \dots, t_s$ be non-negative integers with $t_1\geq 1$. Then, the $\Z_{2^s}$-linear Hadamard code $\bar{H}^{t_1,\dots, t_s}$ belongs to an unique chain of equivalences $C_H(t'_1,\dots,t'_{s'})$, where $s'=s-\sigma+1$.
% If $t_1\geq 2$, then $\sigma=1$ and $\bar{H}^{t_1,\dots,t_s}=C_H(t_1,\dots, t_s)[1]$. Otherwise, if $t_1=1$, then $\sigma>1$ and $\bar{H}^{t_1,\dots, t_s}=C_H(t'_1,\dots,t'_{s'})[\sigma]$, where $(t'_1,\dots,t'_{s'})= (t_\sigma+1,t_{\sigma+1},\dots,t_{s-1},t_s+\sigma-1)$ and $s'=s-\sigma+1$.
\end{proposition}

Now, we recall the results about the rank and dimension of the kernel for nonlinear $\Z_2\Z_4$-linear Hadamard codes, given in \cite{PRV06}.

\begin{theorem}\cite{PRV06}\label{rankZ2Z4}
Let $t\geq 3$ and $t_2 \in \{0,\dots,\lfloor t/2 \rfloor\}$. Let $H^{t_2,t_3}$ be the nonlinear $\Z_2\Z_4$-linear Hadamard code of length $2^t$ and type $(\alpha_1, \alpha_2;t_2,t_3)$, where $\alpha_1=2^{t-t_2}$, $\alpha_2=2^{t-1}-2^{t-t_2-1}$, and $t_3=t+1-2t_2$. Then, $$\rank(H^{t_2,t_3})=t_3+2t_2+ \binom{t_2}{2} \ \mbox{ and } \ \kernel(H^{t_2,t_3})=t_2+t_3.$$
\end{theorem}

The following example shows that the rank, the dimension of the kernel, or the pair of rank and dimension of the kernel can not be used, in general, to completely classify all nonlinear $\Z_2\Z_4\Z_8$-linear Hadamard codes $H^{t_1,t_2,t_3}$ %\cite{Z2Z4Z8Construction} 
of a given length $2^t$.

\begin{example}\label{Example:NoKernel}
By Theorems \ref{Th:H3} and \ref{linearity:Z2Z4Z8}, the nonlinear $\Z_2\Z_4\Z_8$-linear Hadamard codes of length $2^9$ are the following: $H^{1,1,5}$, $H^{1,2,3}$, $H^{1,3,1}$, $H^{2,0,4}$, $H^{2,1,2}$ and $H^{3,0,1}$. Their kernels are of dimension $7,6,5,6,5$ and $4$, respectively, by Theorem \ref{kernel:Z2Z4Z8}. By using the computer algebra system Magma, we can check that $H^{1,2,3}$ and $H^{2,0,4}$ are nonequivalent. Therefore, the dimension of the kernel does not allow to classify these codes. 
By using  Magma, we compute their ranks, which are $12,15,19,15$, $20$ and $26$, respectively. Therefore, the rank does not classify either. Note that $H^{1,2,3}$ and $H^{2,0,4}$ have the same pair of rank and dimension of the kernel, $(r,k)=(6, 15)$.  Thus, the pair of rank and dimension of the kernel does not classify either.
\end{example}

\begin{theorem}\label{rank:t3}
Let $t_1\geq 1$, $t_2\geq 0$, and $t_3\geq 1$ be integers. Let $H^{t_1,t_2,t_3}$ be the $\Z_2\Z_4\Z_8$-linear Hadamard code of type $(\alpha_1,\alpha_2,\alpha_3;t_1,t_2,t_3)$ with $\alpha_1\neq 0$, $\alpha_2\neq 0$ and $\alpha_3\neq 0$.  Then, 
$$
\rank(H^{t_1,t_2,t_3+1})=1+\rank(H^{t_1,t_2,t_3}).
$$
\end{theorem}
\begin{proof}
Let ${\cH}^{t_1,t_2,t_3}=({\cH}_1\mid {\cH}_2\mid {\cH}_3)$ be the corresponding $\Z_2\Z_4Z_8$-additive Hadamard code. Note that, by construction (\ref{eq:recGenMatrix3}),
\begin{align*}
    {\cH}^{t_1,t_2,t_3+1}=\bigcup_{\lambda \in \{0,1\}}(({\cH}_1, {\cH}_1\mid {\cH}_2, {\cH}_2\mid {\cH}_3, {\cH}_3)+\lambda(\zero, \one \mid \zero, \two \mid \zero, \four)).
\end{align*}
By Corollaries \ref{coro4} and  \ref{lemma2}, 
\begin{align*}\label{eq:Ht3p1}
    H^{t_1,t_2,t_3+1}&=\bigcup_{\lambda \in \{0,1\}}(\Phi({\cH}_1, {\cH}_1\mid {\cH}_2, {\cH}_2\mid {\cH}_3, {\cH}_3)+\lambda(\zero, \one,  \zero, \one,  \zero, \one))\nonumber\\
    &=A_0 \cup A_1,
\end{align*}
where $A_\lambda=\Phi({\cH}_1, {\cH}_1\mid {\cH}_2, {\cH}_2\mid {\cH}_3, {\cH}_3)+\lambda(\zero, \one,  \zero, \one,  \zero, \one)$, $\lambda \in \{0,1\}$. Therefore, $\rank(\Phi({\cH}^{t_1,t_2,t_3+1}))=1+\rank(\Phi({\cH}^{t_1,t_2,t_3})).$
\end{proof}

\medskip
Now, we prove some technical lemmas.
\begin{lemma}\label{lem:rnk}
    Let $t$ be a positive integer. Then, 
\begin{enumerate}
    \item $\rank(H^{1,1,t-4})=t+3$, where $t\geq 5$.
    \item $\rank(H^{1,2,t-6})=\rank(H^{2,0,t-5})=t+6$, where $t\geq 7$.
    \item $\rank(H^{1,3,t-8})=t+10$ and $\rank(H^{2,1,t-7})=t+11$, where $t\geq 9$.
     \item $\rank(H^{2,2,t-9})=\rank(H^{3,0,t-8})=t+17$, where $t\geq 10$.
     \item $\rank(H^{1,4,t-10})=t+15$, where $t\geq 11$.
     \item $\rank(\bar{H}^{2,0,0,t-7})=t+7$,  where $t\geq 7$.
     \item $\rank(\bar{H}^{1,1,2,t-10})=t+8$,  where $t\geq 10$.
     \item $\rank(\bar{H}^{1,0,5,t-13})=t+11$,  where $t\geq 13$.
      \item $\rank(\bar{H}^{1,2,0,t-9})=t+9$ and $\rank(\bar{H}^{2,0,1,t-9})=t+11$,  where $t\geq 9$.
     \item $\rank(\bar{H}^{1,1,3,t-12})=t+12$,  where $t\geq 12$.
     \item $\rank(\bar{H}^{1,2,1,t-11})=t+14$,  where $t\geq 11$.
     \item $\rank(\bar{H}^{1,0,6,t-15})=t+16$,  where $t\geq 15$.
\end{enumerate}
\end{lemma}
\begin{proof}
    By using Magma or as it is shown in Tables \ref{table:Types0} and \ref{table:Types1}, we have that $\rank(H^{1,1,1})=8$,  $\rank(H^{1,2,1})=\rank(H^{2,0,2})=13$,  $\rank(H^{1,3,1})=\rank(H^{2,1,1})=19$, $\rank(H^{2,2,1})=\rank(H^{3,0,2})=27$, $\rank(H^{1,4,1})=26$. For any $H^{t_1,t_2,t_3}$, we have that $t+1=3t_1+2t_2+t_3$ by Theorem \ref{Th:H3}. Thus, $H^{1,1,t_3}=H^{1,1,t-4}$, where $t\geq 5$. By Theorem \ref{rank:t3}, we have that $\rank(H^{1,1,t-4})=t+3$. Items $2$-$5$ follow in a similar way. 
    
    Again, by using Magma, we also have that 
    $\rank(\bar{H}^{2,0,0,0})=14$, $\rank(\bar{H}^{1,1,2,0})=18$, $\rank(\bar{H}^{1,2,0,0})=18$, $\rank(\bar{H}^{1,0,5,0})=24$, $\rank(\bar{H}^{2,0,1,0})=20$, $\rank(\bar{H}^{1,1,3,0})=24$, $\rank(\bar{H}^{1,2,1,0})=25$, and $\rank(\bar{H}^{1,0,6,0})=31$.  % Therefore, Items 1 to 5 follow from Theorem \ref{rank:t3}, and 
    The remaining Items $6$-$12$ follow from the fact that $t+1=4t_1+3t_2+2t_3+t_4$ for any $\bar{H}^{t_1,t_2,t_3,t_4}$ and Theorem \ref{rankZ2s}.
\end{proof}

\begin{lemma}\label{lem:equivl}
    Let $t$ be a positive integer.  Then, the following codes in each item are nonequivalent to each other. 
    \begin{enumerate}
        \item $H^{1,2,t-6}$ and $H^{2,0,t-5}$, where $t\geq 7$.
        \item $H^{2,2,t-9}$ and $H^{3,0,t-8}$, where $t\geq 10$.
     \item $H^{1,1,t-4}$ and $\bar{H}^{2,0,t-5}$, where $t\geq 5$.
    \item $H^{2,1,t-7}$ and $\bar{H}^{1,5,t-12}$, where $t\geq 12$.
      \item $H^{2,1,t-7}$ and $\bar{H}^{1,0,5, t-13}$, where $t\geq 13$.
       \item $H^{5,t-9}$ and $H^{2,1,t-7}$, where $t\geq 10$ and $H^{5,t-9}$ is the  $\Z_2\Z_4$-linear Hadamard code of length $2^t$ and type $(\alpha_1, \alpha_2; 5,t-9)$ with $\alpha_1\neq0$ and $\alpha_2\neq0$.
    \end{enumerate}
\end{lemma}
\begin{proof}
Let $H^{5,1}$ be the  $\Z_2\Z_4$-linear Hadamard code of type $(\alpha'_1, \alpha'_2;5,1)$, where $\alpha'_1\neq0$ and $\alpha'_2\neq0$.
    By using Magma, the following codes in each pair are nonequivalent to each other: 
    $(H^{1,2,1}, H^{2,0,2})$, $(H^{2,2,1}, H^{3,0,2})$, $(H^{1,1,1}, \bar{H}^{2,0,0})$, $(H^{2,1,5}, \bar{H}^{1,5,0})$, $(H^{2,1,6}, \bar{H}^{1,0,5,0})$, $(H^{5,1}, H^{2,1,3})$. Therefore, Items $1$ and $2$ follow from construction (\ref{eq:recGenMatrix3}). It is easy to see that Items $3$, $4$ and $5$ follow from constructions (\ref{eq:recGenMatrix3}) and (\ref{eq:recGenMatrix}), and Item $6$ follows from construction (\ref{eq:recGenMatrix3}) and the recursive construction of $\Z_2\Z_4$-linear Hadamard codes given in \cite{RSV08}.
\end{proof}

\begin{lemma}\label{lem:KSig}
    Let $k$ and $t$ be two positive integers such that $t-k>0$. Let $\bar{H}^{t_1,\dots,t_s}$ be a nonlinear $\Z_{2^s}$-linear Hadamard code of length $2^t$ such that $\kernel(\bar{H}^{t_1,\dots,t_s})=t-k$. Then,
    \begin{enumerate}
        \item for $s\ge 2$,
        \begin{equation}\label{EQ:KSig0}
    t_1+\cdots+t_s=t-k-\sigma,
\end{equation}   
\begin{equation}\label{EQ:KSig1}
    (s-1)t_1+ (s-2)t_2+\cdots+t_{s-1}=\sigma+k+1, 
\end{equation}
\item for $s\geq 4$, $t_1=1$, and $\lambda=(s-\sigma-1)t_{\sigma+1}+\cdots+t_{s-1}$,
\begin{equation}\label{EQ:KSig2}
    s-\sigma=\frac{k+2-\lambda}{t_\sigma+1}.
\end{equation}
\end{enumerate}
\end{lemma}
\begin{proof}
   By Theorem \ref{Teo:kernel1}, $\kernel(\bar{H}^{t_1,\dots,t_s})=\sigma+\sum_{i=1}^{s}t_i$, so we have that
\begin{equation}\label{Eq:kr0_}
    t_1+\cdots+t_s=t-k-\sigma.
\end{equation}
By Theorem \ref{Th:Z2Had}, we have that 
\begin{equation}\label{EqHd0_}
    st_1+(s-1)t_2+\cdots+t_s=t+1.
\end{equation}
From (\ref{Eq:kr0_}) and (\ref{EqHd0_}), we have that 
\begin{equation}\label{EqHad_kr0_}
    (s-1)t_1+ (s-2)t_2+\cdots+t_{s-1}=\sigma+k+1.
\end{equation}
 For Item $2$, since $s\geq 4$, $t_1=1$, and $\bar{H}^{t_1,\dots,t_s}$ is nonlinear,  from the definition of $\sigma$, we have that $1<\sigma <s$, $t_2=\cdots=t_{\sigma-1}=0$ and $t_{\sigma}\geq 1$. From (\ref{EqHad_kr0_}), we have that $(s-1)+(s-\sigma)t_{\sigma}+\lambda=\sigma+k+1$. Thus, we obtain (\ref{EQ:KSig2}).
%\begin{equation*}
%    s-\sigma=\frac{k+2-\lambda}{t_\sigma+1}.
%\end{equation*} 
\end{proof}

\begin{lemma}\label{lem:Kt_i}
    Let $k$ and $t$ be two positive integers such that $t-k>0$. Let $H^{t_1,t_2, t_3}$ be a nonlinear $\Z_2\Z_4\Z_8$-linear Hadamard code of length $2^t$ such that $\kernel(H^{t_1,t_2,t_3})=t-k$. Then,
    \begin{equation}\label{EQ:tk0}
   t_1+t_2+t_3=t-k, 
\end{equation}
\begin{equation}\label{EQ:tk1}
   2t_1+t_2=k+1. 
\end{equation}
\end{lemma}
\begin{proof}
    By Corollary \ref{kernel:Z2Z4Z8}, $\kernel(H^{t_1,t_2,t_3})=t_1+t_2+t_3$, so we have that 
\begin{equation}\label{Kr_eq1}
   t_1+t_2+t_3=t-k. 
\end{equation}
By Theorem \ref{Th:H3}, 
\begin{equation}\label{Hda_eq1}
    3t_1+2t_2+t_3=t+1.
\end{equation}
From (\ref{Kr_eq1}) and (\ref{Hda_eq1}), we have that $2t_1+t_2=k+1$. 
\end{proof}

\medskip
The following results show that there are several families of infinite $\Z_2\Z_4\Z_8$-linear Hadamard codes of length $2^t$, 
with $\alpha_1 \neq 0, \alpha_2 \neq 0$ and $\alpha_3 \neq 0$, which are not equivalent to any  $\Z_2\Z_4$-linear and $\Z_{2^s}$-linear Hadamard code, with $s\geq 2$, of the same length $2^t$.

\begin{theorem}\label{TeoClass2} 
The following  $\Z_2\Z_4\Z_8$-linear Hadamard codes  of length $2^t$ are not equivalent to any other $\Z_2\Z_4\Z_8$-linear Hadamard code $H^{t_1,t_2,t_3}$, nor to any $\Z_2\Z_4$-linear or $\Z_{2^s}$-linear Hadamard code, with $s\geq 2$, of length $2^t$:
\begin{enumerate}
    \item $H^{1,1,t-4}$, for $t\geq 5$.
    \item $H^{1,2,t-6}$, for $t\geq 7$, and $H^{2,0,t-5}$, for $t\geq 6$.
    \item $H^{1,3,t-8}$, for $t\geq 9$, and $H^{2,1,t-7}$, for $t\geq 8$.
    \item $H^{1,4,t-10}$, for $t\geq 11$, $H^{2,2,t-9}$ for $t\geq 10$ and $H^{3,0,t-8}$, for $t\geq 9$.
\end{enumerate}
\end{theorem}
\begin{proof}
We consider all $\Z_2\Z_4\Z_8$-linear Hadamard codes $H^{t_1,t_2,t_3}$ of length $2^t$ and dimension of the kernel $t-k$, for $k\in\{2,\dots,5\}$. We will see that, in fact, the codes given in Items $1$-$4$ correspond to the $\Z_2\Z_4\Z_8$-linear Hadamard codes of length $2^t$ and dimension of the kernel $t-k$, for $k\in\{2,\dots,5\}$, respectively. For a given $k$, we define the set 
$$S_k=\{(t_1,t_2,t_3): (t_1,t_2,t_3) \textnormal{ satisfies } (\ref{EQ:tk0}) \textnormal{ and } (\ref{EQ:tk1})\}.$$
By Corollary \ref{kernel:Z2Z4Z8} and Lemma \ref{lem:Kt_i}, we have that $\kernel(H^{t_1,t_2,t_3})=t-k$ if and only if $(t_1,t_2,t_3)\in S_k$. We also define the set
$$\bar{S}_{k,s}=\{(t_1,\dots,t_s): (t_1,\dots,t_s) \textnormal{ satisfies } (\ref{EQ:KSig0}) \textnormal{ and } (\ref{EQ:KSig1})\}.$$ 
By Theorem \ref{Teo:kernel1} and Lemma \ref{lem:KSig}, we have that $\kernel(\bar{H}^{t_1,\dots,t_s})=t-k$ if and only if $(t_1,\dots,t_s)\in \bar{S}_{k,s}$.

\begin{enumerate}
\item Consider $k=2$. We have that $S_2=\{(1,1,t-4)\}$. Let $H_1=H^{1,1,t-4}$. Note that $H_1$ is not equivalent to any other $\Z_2\Z_4\Z_8$-linear Hadamard code $H^{t_1,t_2,t_3}$, having the same length $2^t$, $t\geq 5$, since the values of the kernel dimensions are different. 

Let $H^{t_2,t_3}$ be the nonlinear $\Z_2\Z_4$-linear Hadamard code of length $2^t$ and type $(\alpha_1, \alpha_2;t_2,t_3)$, where $\alpha_1\neq0$ and $\alpha_2\neq0$. By Theorem \ref{rankZ2Z4}, we have that $\rank(H^{t_2,t_3})=t+1+ \binom{t_2}{2}$, and  by Item $1$ of Lemma \ref{lem:rnk}, $\rank(H_1)=t+3$. Since $t+1+\binom{t_2}{2} \neq t+3$ for any value of $t_2$, $H_1$ is not equivalent to any  $\Z_2\Z_4$-linear Hadamard code of length $2^t$ with $\alpha_1\neq0$ and $\alpha_2\neq0$. Since the family of $\Z_4$-linear Hadamard codes is included in the family of $\Z_2\Z_4$-linear Hadamard codes with $\alpha_1\neq0$ and $\alpha_2\neq0$ \cite{KV2015}, we have that $H_1$ is not equivalent to any $\Z_{4}$-linear Hadamard code of length $2^t$.

Now, we only need to show that $H_1$ is not equivalent to any $\Z_{2^s}$-linear Hadamard code, with $s\geq 3$, of length $2^t$. Let $\bar{H}^{t_1,\dots,t_s}$ be a nonlinear $\Z_{2^s}$-linear Hadamard code, with $s\geq 3$, of length $2^t$ such that $\kernel(\bar{H}^{t_1,\dots,t_s})=t-2$. First, we show that, if $s\geq 4$, then $\bar{H}^{t_1,\dots,t_s}$  belongs to a chain of equivalences $C_H(t'_1,\dots,t'_{s'})$, where $s'\in \{2,3\}$. Note that, if $s\geq 4$ and $t_1> 1$, then $\sigma=1$. In this case, (\ref{EQ:KSig1}) has no solution. Assume that $s\geq 4$ and $t_1=1$.  Thus, from (\ref{EQ:KSig2}),
\begin{equation*}\label{Eqlam1}
    s-\sigma=\frac{4-\lambda}{t_\sigma+1}.
\end{equation*}
Since $s-\sigma>0$ and $t_\sigma+1\geq 2$, we have that $4-\lambda \in \{2,3,4\}$, and hence, $s-\sigma \in \{1,2\}$.  Thus, from Proposition \ref{Z2sEqui}, $\bar{H}^{t_1,\dots,t_s}$, with $s\geq 4$,  belongs to a chain of equivalences  $C_H(t'_1,\dots,t'_{s'})$, where $s'\in \{2,3\}$. Therefore, we just need to show that $H_1$ is not equivalent to any nonlinear $\Z_8$-linear Hadamard code $\bar{H}^{t_1,t_2,t_3}$ of length $2^t$ such that $\kernel(\bar{H}^{t_1,t_2,t_3})=t-2$. 

If $t_1=1$, then $\sigma=2$. In this case, we have that $(1,3,t-8)\in \bar{S}_{2,3}$. Otherwise, if $t_1>1$, then $\sigma=1$ and we have that $(2,0,t-5)\in \bar{S}_{2,3}$.  Thus, $\bar{S}_{2,3}= \{(2,0,t-5),(1,3,t-8)\}$. By Theorem \ref{rankZ8}, $\rank(\bar{H}^{2,0,t-5})=t+3$ and $\rank(\bar{H}^{1,3,t-8})=t+4\neq t+3$. From Item $3$ of Lemma \ref{lem:equivl}, $H_1$ and $\bar{H}^{2,0,t-5}$ are nonequivalent to each other. Therefore, $H_1$ is not equivalent to any $\Z_8$-linear Hadamard code of length $2^t$.

\item Consider $k=3$. We have that $S_3=\{(1,2,t-6),(2,0,t-5)\}$. Let $H_1=H^{1,2,t-6}$ and $H_2=H^{2,0,t-5}$. We have that $H_1$ and $H_2$ are not equivalent to any other $\Z_2\Z_4\Z_8$-linear Hadamard code $H^{t_1,t_2,t_3}$, having the same length $2^t$, since the values of the kernel dimensions are different, and $H_1$ and $H_2$ are nonequivalent to each other by Item $1$ of Lemma \ref{lem:equivl}.
    
By Item $2$ of Lemma \ref{lem:rnk}, $\rank(H_1)=\rank(H_2)=t+6$. By using the same argument as in Item $1$, since $t+1+ \binom{t_2}{2}\not=t+6$ for any value of $t_2$,  we have that $H_1$ and $H_2$ are not equivalent to any $\Z_2\Z_4$-linear Hadamard code of length $2^t$ with $\alpha_1\neq0$ and $\alpha_2\neq0$, and neither to any $\Z_{4}$-linear Hadamard code of length $2^t$.

 Now, we show that, if $s\geq 3$, then any $\Z_{2^s}$-linear Hadamard code $\bar{H}^{t_1,\dots,t_s}$ with $\kernel(\bar{H}^{t_1,\dots,t_s})=t-3$ is equivalent to a $\Z_8$-linear Hadamard code. We can assume that $s\geq 4$ and $t_1=1$, otherwise  (\ref{EQ:KSig1}) has no solution. From (\ref{EQ:KSig2}), $5-\lambda \in \{2,3,4,5\}$, and hence, $s-\sigma \in \{1,2\}$. Thus, from Proposition \ref{Z2sEqui}, $\bar{H}^{t_1,\dots,t_s}$, with $s\geq 4$,  belongs to a chain of equivalences $C_H(t'_1,\dots,t'_{s'})$, where $s'\in \{2,3\}$. Therefore,  we just need to show that $H_1$ and $H_2$ are not equivalent to any nonlinear $\Z_8$-linear Hadamard code $\bar{H}^{t_1,t_2,t_3}$ of length $2^t$ such that $\kernel(\bar{H}^{t_1,t_2,t_3})=t-3$. 

By using similar arguments as in Item $1$, $\bar{S}_{3,3}=\{(2,1,t-7),(1,4,t-10)\}$. By Theorem \ref{rankZ8}, $\rank(\bar{H}^{2,1,t-7})=t+5$, and $\rank(\bar{H}^{1,4,t-10})=t+7$. This completes the proof, since $\rank(H_1)=\rank(H_2)=t+6$ by Item $2$ of Lemma \ref{lem:rnk}.

\item Consider $k=4$. We have that $S_4=\{(1,3,t-8),(2,1,t-7)\}$. Let $H_1=H^{1,3,t-8}$ and $H_2=H^{2,1,t-7}$. We have that $H_1$ and $H_2$ are not equivalent to any other $\Z_2\Z_4\Z_8$-linear Hadamard code $H^{t_1,t_2,t_3}$, having the same length $2^t$, since the values of the kernel dimensions are different. Moreover, from Item $3$ of Lemma \ref{lem:rnk}, $\rank(H_1)=t+10$ and $\rank(H_2)=t+11$, and hence $H_1$ and $H_2$ are nonequivalent to each other. 

 By using the same argument as in Item $1$, since $t+1+ \binom{t_2}{2}\not=t+10$ for any value of $t_2$, we have that $H_1$ is not equivalent to any $\Z_2\Z_4$-linear Hadamard code of length $2^t$ with $\alpha_1\neq0$ and $\alpha_2\neq0$. On the other hand, if $t+1+ \binom{t_2}{2}=t+11$, then $t_2=5$, and $t_3=t-9$ by Theorem \ref{rankZ2Z4}. Let $H^{5,t-9}$ be the  $\Z_2\Z_4$-linear Hadamard code of length $2^t$ and type $(\alpha_1, \alpha_2; 5,t-9)$, with $\alpha_1\neq0$ and $\alpha_2\neq 0$. From Item $6$ of Lemma \ref{lem:equivl},  $H^{5,t-9}$ and $H_2$ are nonequivalent to each other. Thus, $H_1$ and $H_2$ are not equivalent to any  $\Z_2\Z_4$-linear Hadamard code of length $2^t$ with $\alpha_1\neq0$ and $\alpha_2\neq0$, and neither to any $\Z_{4}$-linear Hadamard code of length $2^t$. 

 Now, we show that any $\Z_{2^s}$-linear Hadamard code  $\bar{H}^{t_1,\dots,t_s}$ with $s\geq 3$ and $\kernel(\bar{H}^{t_1,\dots,t_s})=t-4$  is equivalent to a $\Z_8$-linear or  $\Z_{16}$-linear Hadamard code. For $s\geq4$ and $t_1>1$, we have that $\sigma=1$, so (\ref{EQ:KSig1}) has solution  only if $s=4$ and hence $s-\sigma=3$. For $s\geq4$ and $t_1=1$, from (\ref{EQ:KSig2}), we have that $6-\lambda \in \{2,3,4,5,6\}$, and hence $s-\sigma \in \{1,2,3\}$.  Thus, from Proposition \ref{Z2sEqui}, $\bar{H}^{t_1,\dots,t_s}$, with $s\geq 4$,  belongs to a chain of equivalences $C_H(t'_1,\dots,t'_{s'})$, where $s'\in \{2,3,4\}$. Therefore, we just need to show that $H_1$ and $H_2$ are not equivalent to any nonlinear $\Z_8$-linear and $\Z_{16}$-linear Hadamard code of length $2^t$.

We have that 
\begin{align*}
\bar{S}_{4,3}\cup \bar{S}_{4,4}=\{ &(1,5,t-12),(2,2,t-9), (3,0,t-8), (1,0,5, t-13),\\
             & (1,1,2, t-10), (1,2,0,t-9),(2,0,0, t-7)\}.
\end{align*}
 By Theorem \ref{rankZ8} and Items $6$-$9$ of Lemma \ref{lem:rnk}, we have that $\rank(\bar{H}^{t_1,\dots,t_s})$ is $t+11, t+8,t+9,t+11,t+8,t+9$ and $t+7$, respectively. Thus, we only have to consider $H_2$, $\bar{H}^{1,5,t-12}$, and $\bar{H}^{1,0,5,t-13}$, because they have the same rank. From Item $4$ of Lemma \ref{lem:equivl}, $H_2$ and $\bar{H}^{1,5,t-12}$ are nonequivalent to each other. Moreover, from Item $5$ of Lemma \ref{lem:equivl}, $H_2$ and $\bar{H}^{1,0,5,t-13}$ are nonequivalent to each other. Therefore, $H_1$ and $H_2$ are not equivalent to any $\Z_{2^s}$-linear Hadamard code $\bar{H}^{t_1,\ldots,t_s}$ of length $2^t$, $s\geq 3$.

\item Finally, consider $k=5$.  We have that $$S_5=\{(1,4,t-10), (2,2,t-9), (3,0,t-8)\}.$$ Let $H_1=H^{1,4,t-10}$, $H_2=H^{2,2,t-9}$ and $H_3=H^{3,0,t-8}$. By Items $4$ and $5$ of Lemma \ref{lem:rnk}, $\rank(H_1)=t+15$ and $\rank(H_2)=\rank(H_3)=t+17$. Moreover, by Item $2$ of Lemma \ref{lem:equivl}, $H_2$ and $H_3$ are nonequivalent to each other. Therefore, $H_1$, $H_2$, and $H_3$ are not equivalent to any other $\Z_2\Z_4\Z_8$-linear Hadamard code $H^{t_1,t_2,t_3}$, having the same length $2^t$, and they are nonequivalent to each other. 

By using the same argument as in Item $1$, since $t+1+ \binom{t_2}{2}$ is not equal to $t+15$ or $t+17$, we have that $H_1, H_2$ and $H_3$ are not equivalent to any  $\Z_2\Z_4$-linear Hadamard code of length $2^t$ with $\alpha_1\neq0$ and $\alpha_2\neq0$, and neither to any $\Z_{4}$-linear Hadamard code of length $2^t$. 

 Now, we show that any $\Z_{2^s}$-linear Hadamard code $\bar{H}^{t_1,\dots,t_s}$ with $s\geq 3$ and $\kernel(\bar{H}^{t_1,\dots,t_s})=t-5$ is equivalent to a $\Z_8$-linear or $\Z_{16}$-linear Hadamard code. For $s\geq 4$ and $t_1>1$, we have that $\sigma=1$, so  (\ref{EQ:KSig1}) has a solution only if $s=4$ and hence $s-\sigma=3$. For $s\geq4$ and $t_1=1$, from (\ref{EQ:KSig2}), we have that $7-\lambda \in \{2,3,4,5,6,7\}$, and hence $s-\sigma \in \{1,2,3\}$.  Thus, from Proposition \ref{Z2sEqui}, $\bar{H}^{t_1,\dots,t_s}$, with $s\geq 4$,  belongs to a chain of equivalences $C_H(t'_1,\dots,t'_{s'})$, where $s'\in \{2,3,4\}$. Therefore, we just need to show that $H_1$, $H_2$, and $H_3$ are not equivalent to any nonlinear $\Z_8$-linear and $\Z_{16}$-linear Hadamard code of length $2^t$.

We have that 
\begin{align*}
\bar{S}_{5,3}\cup \bar{S}_{5,4}=\{ & (1,6,t-14),(2,3,t-11), (3,1,t-10),(1,0,6, t-15),\\ 
             & (1,1,3, t-12), (1,2,1,t-11),(2,0,1, t-9)\}. 
\end{align*}
By Theorem \ref{rankZ8} and from Items $9$ to $12$ of Lemma \ref{lem:rnk}, we have that $\rank(\bar{H}^{t_1,\dots,t_s})$ is $t+16, t+12,t+14,t+16,t+12,t+14$ and $t+11$, respectively. Therefore, $H_1, H_2$ and $H_3$ are not equivalent to any $\Z_{2^s}$-linear Hadamard code of length $2^t$, $s\geq 3$.
\end{enumerate}
\end{proof}

\begin{theorem}
The  $\Z_2\Z_4\Z_8$-linear Hadamard codes $H^{t/3,0,1}$,  with $t=3m$ and $m\geq 2$,  of length $2^t$ are not equivalent to any other $\Z_2\Z_4\Z_8$-linear Hadamard code $H^{t_1,t_2,t_3}$, nor to any $\Z_2\Z_4$-linear, $\Z_4$-linear or $\Z_8$-linear Hadamard code of length $2^t$.  
\end{theorem}
\begin{proof}
By Theorem \ref{linearity:Z2Z4Z8}, $H^{m,0,1}$ is nonlinear since $m\geq 2$. Then, by Corollary \ref{kernel:Z2Z4Z8},  $\kernel(H^{m,0,1})=m+1$.
Let $H^{t_1,t_2, t_3}$ be a nonlinear $\Z_2\Z_4\Z_8$-linear Hadamard code of length $2^t$ such that $\kernel(H^{t_1,t_2,t_3})=m+1$.  By Corollary \ref{kernel:Z2Z4Z8}, $\kernel(H^{t_1,t_2,t_3})=t_1+t_2+t_3$, so we have that 
\begin{equation}\label{Ker_eq0_}
   t_1+t_2+t_3=m+1. 
\end{equation}
By Theorem \ref{Th:H3}, 
\begin{equation}\label{Hada_eq0_}
    3t_1+2t_2+t_3=t+1=3m+1.
\end{equation}
From (\ref{Ker_eq0_}) and (\ref{Hada_eq0_}),
\begin{equation}\label{HadaKer_eq0_1}
   t_2+2t_3=2. 
\end{equation}
Note that $\{ (t_1,t_2,t_3) : (t_1,t_2,t_3) \mbox{ satisfies } (\ref{Ker_eq0_})\mbox{ and } (\ref{HadaKer_eq0_1}) \}=\{(m,0,1)\}$.
Therefore, $H^{m,0,1}$ is not equivalent to any other $\Z_2\Z_4\Z_8$-linear Hadamard code $H^{t_1,t_2,t_3}$ of the same length $2^t$, since the values of the kernel dimensions are different.

Let $H^{t_2,t_3}$ be the nonlinear  $\Z_2\Z_4$-linear Hadamard code of length $2^t$ and type $(\alpha_1, \alpha_2; t_2,t_3)$, where $\alpha_1\neq0$ and $\alpha_2\neq0$, such that $\kernel(H^{t_2,t_3})=m+1$. Then, by Theorem \ref{rankZ2Z4}, 
\begin{equation}\label{EqC0}
    t_2+t_3=m+1,
\end{equation}
and 
\begin{equation}\label{EqC1}
    2t_2+t_3=t+1=3m+1.
\end{equation}
From (\ref{EqC0}) and (\ref{EqC1}), 
\begin{equation}\label{EqC2}
    t_2+2t_3=2.
\end{equation}
From (\ref{EqC0}) and (\ref{EqC2}), we have that $t_3=1-m<0$, which is a contradiction, since $t_3\geq 1$.  Therefore, $H^{m,0,1}$ is not equivalent to any  $\Z_2\Z_4$-linear Hadamard code of length $2^t$ with $\alpha_1\neq0$ and $\alpha_2\neq0$. Since the family of $\Z_4$-linear Hadamard codes is included in the family of $\Z_2\Z_4$-linear Hadamard codes with $\alpha_1\neq0$ and $\alpha_2\neq0$, we only need to show that $H^{m,0,1}$ is not equivalent to any $\Z_{8}$-linear Hadamard code of length $2^t$. 

Let $\bar{H}^{t_1,t_2,t_3}$ be a nonlinear $\Z_{8}$-linear Hadamard code of length $2^t$ such that $\kernel(\bar{H}^{t_1,t_2,t_3})=m+1$. By Theorem \ref{Teo:kernel1}, $\kernel(\bar{H}^{t_1,t_2,t_3})=\sigma+\sum_{i=1}^{3}t_i$, so we have that
\begin{equation}\label{Eq:kereq0_}
    t_1+t_2+t_3=m+1-\sigma.
\end{equation}
By Theorem \ref{Th:Z2Had}, we have that 
\begin{equation}\label{EqHad0_}
    3t_1+2t_2+t_3=t+1=3m+1.
\end{equation}
From (\ref{Eq:kereq0_}) and (\ref{EqHad0_}), we have that 
\begin{equation}\label{EqHad_ker0_}
    t_2+2t_3=2-3\sigma.
\end{equation}
Since $1\leq \sigma\leq 3$, $2-3\sigma<0$. Thus, (\ref{EqHad_ker0_}) has no solution, since $t_2,t_3\geq 0$. Therefore, $H^{m,0,1}$ is not equivalent to any  $\Z_8$-linear Hadamard code.
\end{proof}

\begin{theorem}
   The  $\Z_2\Z_4\Z_8$-linear Hadamard codes  $H^{(t-2)/3,1,1}$,  with $t=3m+2$ and $m\geq 2$,  of length $2^t$ are not equivalent to any other $\Z_2\Z_4\Z_8$-linear Hadamard code $H^{t_1,t_2,t_3}$, nor to any $\Z_4$-linear or $\Z_2\Z_4$-linear Hadamard code of length $2^t$.   
\end{theorem}
\begin{proof}
By Theorem \ref{linearity:Z2Z4Z8}, $H^{m,1,1}$ is nonlinear. Then, by Corollary \ref{kernel:Z2Z4Z8}, we have that $\kernel(H^{m,1,1})=m+2$.
    Let $H^{t_1,t_2, t_3}$ be a nonlinear $\Z_2\Z_4\Z_8$-linear Hadamard code of length $2^t$ such that $\kernel(H^{t_1,t_2,t_3})=m+2$.  By Corollary \ref{kernel:Z2Z4Z8}, $\kernel(H^{t_1,t_2,t_3})=t_1+t_2+t_3$, so we have that 
\begin{equation}\label{Ker_eq0__}
   t_1+t_2+t_3=m+2. 
\end{equation}
By Theorem \ref{Th:H3}, 
\begin{equation}\label{Hada_eq0__}
    3t_1+2t_2+t_3=t+1=3m+3.
\end{equation}
From (\ref{Ker_eq0__}) and (\ref{Hada_eq0__}),
\begin{equation}\label{HadaKer_eq0__}
   t_2+2t_3=3. 
\end{equation}
Note that $\{(t_1,t_2,t_3): (t_1,t_2,t_3) \textnormal{ satisfies } (\ref{Ker_eq0__}) \textnormal{ and } (\ref{HadaKer_eq0__})\}=\{(m,1,1)\}$.
Therefore, $H^{m,1,1}$ is not equivalent to any other $\Z_2\Z_4\Z_8$-linear Hadamard code $H^{t_1,t_2,t_3}$, having the same length $2^t$, since the values of the kernel dimensions are different.

Let $H^{t_2,t_3}$ be the nonlinear  $\Z_2\Z_4$-linear Hadamard code of length $2^t$ and type $(\alpha_1, \alpha_2; t_2,t_3)$, where $\alpha_1\neq0$ and $\alpha_2\neq0$, such that $\kernel(H^{t_2,t_3})=m+2$. Then, by Theorem \ref{rankZ2Z4}, 
\begin{equation}\label{EqC0_}
    t_2+t_3=m+2,
\end{equation}
and 
\begin{equation}\label{EqC1_}
    2t_2+t_3=t+1=3m+3.
\end{equation}
From (\ref{EqC0_}) and (\ref{EqC1_}), 
\begin{equation}\label{EqC2_}
    t_2+2t_3=3.
\end{equation}
From (\ref{EqC0_}) and (\ref{EqC2_}), we have that $t_3=1-m<0$, which is a contradiction, since $t_3\geq 1$.  Therefore, $H^{m,1,1}$ is not equivalent to any  $\Z_2\Z_4$-linear Hadamard code of length $2^t$ with $\alpha_1\neq0$ and $\alpha_2\neq0$. Since the family of $\Z_4$-linear Hadamard codes is included in the family of $\Z_2\Z_4$-linear Hadamard codes with $\alpha_1\neq0$ and $\alpha_2\neq0$. This completes the proof. 
\end{proof}

\medskip
For $4\leq t\leq 15$, Tables \ref{table:Types0}, \ref{table:Types1} and \ref{table:Types2},  show all possible %values of ($t_1,t_2,t_3$)
types corresponding to $\Z_4$-linear Hadamard, $\Z_2\Z_4$-linear Hadamard (with $\alpha_1 \neq 0$ and $\alpha_2\neq 0$), $\Z_8$-linear Hadamard $\bar{H}^{t_1,t_2,t_3}$ \cite{KernelZ2s} and $\Z_2\Z_4\Z_8$-linear (with $\alpha_1 \neq 0$, $\alpha_2 \neq 0$ and $\alpha_3\neq 0$) Hadamard codes $H^{t_1,t_2,t_3}$ %\cite{Z2Z4Z8Construction}  
of length $2^t$. For each one of them, the values $(r,k)$ are shown, where $r$ is the rank and $k$ is the dimension of the kernel. These values can be computed for $\Z_4$-linear Hadamard codes from \cite{Kro:2001:Z4_Had_Perf, PRV06}, for $\Z_2\Z_4$-linear Hadamard codes  by Theorem \ref{rankZ2Z4}, and for $\Z_8$-linear Hadamard codes by Theorems \ref{Teo:kernel1} and \ref{rankZ8}.
The values of the rank for $\Z_2\Z_4\Z_8$-linear Hadamard codes are computed by using Magma, and the kernel dimensions are calculated by Theorem \ref{kernel:Z2Z4Z8}. Note that if two codes have different values $(r,k)$, then they are not equivalent. As shown in Example \ref{Example:NoKernel} for $t=9$, it is easy to see that taking only the values of the dimension of the kernel, or only the rank, given in these tables, it is not possible to classify completely the $\Z_2\Z_4\Z_8$-linear Hadamard codes. 
%Also, considering only the rank, it is not possible to fully classify these codes either. 

Note that, for $t\leq 4$, the $\Z_2\Z_4\Z_8$-linear Hadamard codes $H^{t_1,t_2,t_3}$ %\cite{Z2Z4Z8Construction}, 
%with $\alpha_1 \neq 0$, $\alpha_2 \neq 0$ and $\alpha_3\neq 0$, 
of length $2^t$ are binary linear Hadamard codes.
In the next theorem, we show that, for $5\leq t\leq 11$,  the nonlinear $\Z_2\Z_4\Z_8$-linear Hadamard codes $H^{t_1,t_2,t_3}$ %\cite{Z2Z4Z8Construction}
%with $\alpha_1 \neq 0$, $\alpha_2 \neq 0$ and $\alpha_3\neq 0$  
of length $2^t$ are not equivalent to each other, 
%to any  $\Z_2\Z_4\Z_8$-linear Hadamard code $H^{t_1,t_2,t_3}$ % \cite{Z2Z4Z8Construction} 
%of any other type, 
nor to any $\Z_2\Z_4$-linear Hadamard code, nor to any  $\Z_{2^s}$-linear Hadamard code \cite{KernelZ2s} with $s\geq 2$, of the same length $2^t$.
 
\begin{theorem}\label{Th:5_t_11}
For $5\leq t\leq 11$, the nonlinear $\Z_2\Z_4\Z_8$-linear Hadamard codes $H^{t_1,t_2,t_3}$  of length $2^t$ are not equivalent to each other, nor to any  $\Z_2\Z_4$-linear Hadamard code, nor to any previously constructed $\Z_{2^s}$-linear Hadamard code, with $s\geq 2$, of length $2^t$.
\end{theorem}
\begin{proof}
It follows from %Proposition \ref{H1H2Non},  Theorems \ref{TeoClass1} and \ref{TeoClass2},
Theorem \ref{TeoClass2}, Tables \ref{table:Types0} and \ref{table:Types1}, Tables 1 and 3 given in \cite{KernelZ2s}, and the fact that two codes are nonequivalent if they have different values of $(r,k)$, where $r$ is the rank and $k$ is the dimension of the kernel. For example, consider $t=11$. From Table \ref{table:Types1}, all nonlinear $\Z_2\Z_4\Z_8$-linear Hadamard codes  of length $2^{11}$ are $H^{1,1,7}$, $H^{1,2,5}$, $H^{1,3,3}$, $H^{1,4,1}$, $H^{2,0,6}$, $H^{2,1,4}$, $H^{2,2,2}$, $H^{3,0,3}$ and $H^{3,1,1}$. All these codes, except $H^{2,2,2}$ and $H^{3,0,3}$,  are  not equivalent to each other, nor to any $\Z_2\Z_4$-linear and $\Z_{2^s}$-linear Hadamard code, with $s\geq 2$, of length $2^{11}$, since they have different values of $(r,k)$. The codes $H^{2,2,2}$ and $H^{3,0,3}$ have the same $(r,k)$, which is $(28,6)$. By using Magma, we can check that they are not equivalent. Again, from Table \ref{table:Types1}, and Table 3 given in \cite{KernelZ2s}, we have that all the codes of length $2^{11}$ have values of $(r,k)$ different from $(28,6)$. This completes the proof.
\end{proof}

Let $\mathcal{A}_t$ be the number of nonequivalent $\Z_2\Z_4\Z_8$-linear Hadamard codes $H^{t_1,t_2,t_3}$  of length $2^t$. By Theorem \ref{Th:5_t_11}, we can classify completely $\Z_2\Z_4\Z_8$-linear Hadamard codes $H^{t_1,t_2,t_3}$ of length $2^t$ for $3\leq t\leq 11$, by providing the number $\mathcal{A}_t$ of nonequivalent such codes,  as shown in Table \ref{table:ClassificationEquiv}.
\begin{table}[h]
\centering
\begin{tabular}{|c||c|c|c|c|c|c|c|c|c|}
\cline{0-9}
$t$ & 3 & 4 & 5 & 6 & 7 & 8 & 9 & 10 &11 \\ \cline{0-9}
$\mathcal{A}_t$ &  1 &  1 &  2 &  3 &  4 &  5 &  7   &8 &10\\\cline{0-9}
\end{tabular}
\caption{Number $\mathcal{A}_t$ of nonequivalent $\Z_2\Z_4\Z_8$-linear Hadamrd codes of length $2^t$ for $3\leq t\leq 11$.}
\label{table:ClassificationEquiv}
\end{table}

\section{Conclusions and further research}
\label{sec:conclusions}

In this paper, we study the linearity of the $\Z_2\Z_4\Z_8$-linear Hadamard codes $H^{t_1,t_2,t_3}$ 
with $\alpha_1\neq0$, $\alpha_2\neq0$, and $\alpha_3\neq0$, presented in \cite{Z2Z4Z8Construction}. We also determine the kernel and its dimension whenever they are nonlinear. The kernel does not give a complete classification, but it provides a lower bound on the number of such nonequivalent $\Z_2\Z_4\Z_8$-linear Hadamard codes with the same length $2^t$. %Then, the rank of some of these codes has been computed by using  Magma. 
We establish a complete classification for $3\leq t\leq 11$ by using the dimension of the kernel and some Magma computations. It is known that  each $\Z_4$-linear Hadamard code is equivalent to a $\Z_2\Z_4$-linear Hadamard codes with $\alpha_1\neq 0$ and $\alpha_2\neq 0$ \cite{KV2015}. Unlike $\Z_2\Z_4$-linear Hadamard codes, in general, the family of $\Z_2\Z_4\Z_8$-linear Hadamard codes $H^{t_1,t_2,t_3}$ does not include the family of $\Z_4$-linear or $\Z_8$-linear Hadamard codes. Actually, we show that there are several families of infinite nonlinear $\Z_2\Z_4\Z_8$-linear Hadamard codes 
with $\alpha_1\neq0$, $\alpha_2\neq0$, and $\alpha_3\neq0$, which are not equivalent to any other $\Z_2\Z_4\Z_8$-linear Hadamard code $H^{t_1,t_2,t_3}$, nor to any  $\Z_2\Z_4$-linear Hadamard, nor to any $\Z_{2^s}$-linear Hadamard code \cite{KernelZ2s} with $s\geq 2$, of the
 same length $2^t$. Therefore, some nonlinear Hadamard codes, without any known structure, now can be seen as the Gray map image of a $\Z_2\Z_4\Z_8$-additive Hadamard code with $\alpha_1\neq0$, $\alpha_2\neq0$, and $\alpha_3\neq0$. 

 As further research, it would be interesting to prove that all nonlinear $\Z_2\Z_4\Z_8$-linear Hadamard codes $H^{t_1,t_2,t_3}$ are not equivalent to each other, nor to  any  $\Z_2\Z_4$-linear Hadamard, nor to any  $\Z_{2^s}$-linear Hadamard code \cite{KernelZ2s} with $s\geq 2$, of the
 same length $2^t$, as it is seen for $5\leq t\leq 11$ by Theorem \ref{Th:5_t_11}.

%\begin{table}[h!]
\begin{table}[!ht]
\footnotesize
\centering
%\begin{tabular}{|c||cc|cc|cc|cc|}
\begin{tabular}{|p{0.8cm}| p{1.5cm}p{0.7cm} | p{1.75cm}p{0.7cm} | p{1.85cm}p{0.7cm} | p{1.85cm}p{0.7cm} |}
\cline{1-9}
& \multicolumn{2}{c|}{$t=4$}& \multicolumn{2}{c|}{$t=5$}& \multicolumn{2}{c|}{$t=6$}& \multicolumn{2}{c|}{$t=7$}\\
\cline{2-9}
& type & $(r,k)$ & type & $(r,k)$ & type & $(r,k)$ & type & $(r,k)$ \\[0.2em]
\hline
\multirow{2}{*}{$\Z_4$}&$(2^3; 1,3)$ &(5,5) &$(2^4;1,4)$ &(6,6) &$(2^5;1,5)$ &(7,7) &$(2^6;1,6)$ &(8,8) \\
&$(2^3; 2,1)$ &(5,5) &$(2^4;2,2)$ &(6,6) &$(2^5;2,3)$ &(7,7) &$(2^6;2,4)$  &(8,8)\\
&  &  &$(2^4;3,0)$  & (7,4)& $(2^5;3,1)$  & (8,5)  & $(2^6;3,2)$  & (9,6) \\
                       & & &          &      &          &        & $(2^6;4,0)$  & (11,5)\\[0.2em]
\hline

\multirow{4}{*}{$\Z_2\Z_4$} &$(8,4;1,3)$   &(5,5) & $(16,8;1,4)$  &(6,6)      &$(32,16;1,5)$             &(7,7)    &$(64,32;1,6)$  &(8,8)   \\
                             %&$(16,0;0,5)$ &(5,5)  & $(32,0;0,6)$  &(6,6)       &$(64,0;0,7)$             &(7,7)        &$(128,0;0,8)$  &(8,8)   \\
                         &$(4,6;2,1)$  &(6,3) & $(8,12;2,2)$  &(7,4)        &$(16,24;2,3)$  &(8,5)       &$(32,48;2,4)$  &(9,6)   \\
                           &&   & &            &$(8,28;3,1)$       &(10,4)             &$(16,56;3,2)$        &(11,5)              \\[0.2em]
\hline
\multirow{5}{*}{$\Z_8$}&$(2^2; 1,0,2)$ &(5,5) &$(2^3;1,0,3)$ &(6,6) &$(2^4;1,0,4)$ &(7,7) &$(2^5;1,0,5)$   &(8,8)\\
&$(2^2; 1,1,0)$ &(5,5) &$(2^3; 1,1,1)$ &(6,6) &$(2^4;1,1,2)$  &(7,7) &$(2^5;1,1,3)$ &(8,8)\\
& & & $(2^3;2,0,0)$ & (8,3) & $(2^4;1,2,0)$ & (8,5)  & $(2^5;1,2,1)$ & (9,6)  \\
                       & &&           &       & $(2^4;2,0,1)$ & (9,4)  & $(2^5;2,0,2)$ & (10,5)  \\
                        & & &          &       &           &        & $(2^5;2,1,0)$ & (12,4)  \\[0.2em]
\hline
\multirow{3}{*}{$\Z_2\Z_4\Z_8$} &$(4,2,2;1,0,2)$   &(5,5)  &$(8,4,4;1,0,3)$ &(6,6)       &$(16,8,8;1,0,4)$   &(7,7)        &$(32,16,16;1,0,5)$  &(8,8)   \\
                           %&$(16,0,0;0,0,5)$&(5,5)  &$(32,0,0;0,0,6)$&(6,6)  &$(64,0,0;0,0,7)$ &(7,7)            &$(128,0,0;0,0,8)$               &(8,8)         \\
                           & &   & $(4,6,4;1,1,1)$ &(8,3)    & $(8,12,8;1,1,2)$  &(9,4)         &$(16,24,16;1,1,3)$               &(10,5)         \\
                           && &&  & $(4,6,12;2,0,1)$   &(12,3) &$(8,28,16;1,2,1)$&(13,4) \\
                           && && && &$(8,12,24;2,0,2)$  &(13,4)\\[0.2em]
\hline
\end{tabular}
\caption{Rank and kernel for all types of  $\Z_4$-linear, $\Z_8$-linear, $\Z_2\Z_4$-linear, and $\Z_2\Z_4\Z_8$-linear Hadamard codes of length $2^t$, $4\leq t\leq 7$.}
\label{table:Types0}
\end{table}

%\begin{table}[h!]
\begin{table}[!ht]
\tiny
\centering
%\begin{tabular}{|c||cc|cc|cc|cc|}
\begin{tabular}{|p{0.6cm}| p{1.5cm}p{0.7cm} | p{1.75cm}p{0.7cm} | p{1.85cm}p{0.7cm} | p{1.85cm}p{0.7cm} |}
\cline{1-9}
& \multicolumn{2}{c|}{$t=8$}& \multicolumn{2}{c|}{$t=9$}& \multicolumn{2}{c|}{$t=10$}& \multicolumn{2}{c|}{$t=11$}\\
\cline{2-9}
& type & $(r,k)$ & type & $(r,k)$ & type & $(r,k)$ & type & $(r,k)$ \\[0.2em]
\hline
\multirow{2}{*}{$\Z_4$}
&$(2^7;1,7)$ &(9,9) &$(2^8;1,8)$ &(10,10) &$(2^9;1,9)$ &(11,11) &$(2^{10};1,10)$ &(12,12) \\

&$(2^7;2,5)$ &(9,9) &$(2^8;2,6)$ &(10,10) &$(2^9;2,7)$ &(11,11) &$(2^{10};2,8)$  &(12,12)\\

 &$(2^7;3,3)$ &(10,7)& $(2^8;3,4)$  & (11,8)  & $(2^9;3,5)$  & (12,9) &$(2^{10};3,6)$ &(13,10) \\
  
&$(2^7;4,1)$ &(12,6) &$(2^8;4,2)$          &(13,7)      &$(2^9;4,3)$          &(14,8)        & $(2^{10};4,4)$  & (15,9)\\

&& &$(2^8;5,0)$ &(16,6) &$(2^9;5,1)$ &(17,7) &$(2^{10};5,2)$ &(18,8)\\

&& && && &$(2^{10};6,0)$  &(22,7) \\[0.2em]

\hline
\multirow{4}{*}{$\Z_2\Z_4$} &$(128,64;1,7)$ &(9,9) &$(256,128;1,8)$ &(10,10) &$(512,256;1,9)$ &(11,11) &$(1024,512;1,10)$ &(12,12) \\

%&$(256,0;0,9)$&(9,9) &$(512,0;0,10)$&(10,10) &$(1024,0;0,11)$&(11,11) &$(2048,0;0,12)$&(12,12)\\

&$(64,96;2,5)$ &(10,7) &$(128,192;2,6)$ &(11,8) &$(256,384;2,7)$ &(12,9) &$(512,768;2,8)$  &(13,10)\\

  &$(32,112;3,3)$  & (12,6)& $(64,224;3,4)$  & (13,7)  & $(128,448;3,5)$  & (14,8) &$(256,896;3,6)$ &(15,9) \\
  
&$(16,120;4,1)$ &(15,5) &$(32,240;4,2)$          &(16,6)      &$(64,480;4,3)$          &(17,7)        & $(128,960;4,4)$  & (18,8)\\

&& && &$(32,496;5,1)$ &(21,6) &$(64,992;5,2)$ &(22,7)\\[0.2em]

\hline
\multirow{5}{*}{$\Z_8$}&$(2^6;1,0,6)$ &(9,9) &$(2^7;1,0,7)$ &(10,10) &$(2^8;1,0,8)$ &(11,11) &$(2^9;1,0,9)$   &(12,12)\\

&$(2^6;1,1,4)$ &(9,9) &$(2^7;1,1,5)$ &(10,10) &$(2^8;1,1,6)$  &(11,11) &$(2^9;1,1,7)$ &(12,12)\\

&$(2^6;1,2,2)$ &(10,7) &$(2^7;1,2,3)$ & (11,8) & $(2^8;1,2,4)$ & (12,9)  & $(2^9;1,2,5)$ & (13,10)  \\

&$(2^6;1,3,0)$ &(12,6) &$(2^7;1,3,1)$  &(13,7)       & $(2^8;1,3,2)$ & (14,8)  & $(2^9;1,3,3)$ & (15,9)  \\

&$(2^6;2,0,3)$ &(11,6) &$(2^7;2,0,4)$          &(12,7)       &$(2^8;1,4,0)$  &(17,7)        & $(2^9;1,4,1)$ & (18,8)  \\

&$(2^6;2,1,1)$ &(13,5) &$(2^7;2,1,2)$  &(14,6) &$(2^8;2,0,5)$ &(13,8) &$(2^9;2,0,6)$ &(14,9)\\

&$(2^6;3,0,0)$  &(17,4) &$(2^7;2,2,0)$ &(17,5) &$(2^8;2,1,3)$ &(15,7) &$(2^9;2,1,4)$    &(16,8) \\

&& &$(2^7;3,0,1)$ &(18,5) &$(2^8;2,2,1)$ &(18,6) &$(2^9;2,2,2)$ &(19,7) \\

&& && &$(2^8;3,0,2)$ &(19,6) &$(2^9;2,3,0)$ &(23,6)\\

&& && &$(2^8;3,1,0)$ &(24,5) &$(2^9;3,0,3)$ &(20,7)\\

 && && && &$(2^9;3,1,1)$ &(25,6)\\
 
 && && && &$(2^9;4,0,0)$ &(32,5)\\[0.2em]
\hline

\multirow{3}{*}{$\Z_2\Z_4\Z_8$} &$(64,32,32;1,0,6)$ &(9,9) &$(128,64,64;1,0,7)$ &(10,10) &$(256,128,128;1,0,8)$ &(11,11) &$(512,256,256;1,0,9)$   &(12,12)\\

%&$(256,0,0;0,0,9)$&(9,9) &$(512,0,0;0,0,10)$&(10,10) &$(1024,0,0;0,0,11)$&(11,11) &$(2048,0,0;0,0,12)$&(12,12)\\

&$(32,48,32;1,1,4)$ &(11,6) &$(64,96,64;1,1,5)$ &(12,7) &$(128,192,128;1,1,6)$  &(13,8) &$(256,384,256;1,1,7)$ &(14,9)\\

&$(16,56,32;1,2,2)$ &(14,5) &$(32,112,64;1,2,3)$ & (15,6) & $(64,224,128;1,2,4)$ & (16,7)  & $(128,448,256;1,2,5)$ & (17,8)  \\

&$(16,24,48;2,0,3)$ &(14,5) &$(16,120,64;1,3,1)$  &(19,5)       & $(32,240,128;1,3,2)$ & (20,6)  & $(64,480,256;1,3,3)$ & (21,7)  \\

&$(8,28,48;2,1,1)$ &(19,4) &$(32,48,96;2,0,4)$   &(15,6)       &$(64,96,192;2,0,5)$ &(16,7)        & $(32,496,256;1,4,1)$ & (26,6)  \\

&& &$(16,56,96;2,1,2)$  &(20,5) &$(32,112,192;2,1,3)$ &(21,6) &$(128,192,384;2,0,6)$ &(17,8)\\

&& &$(8,28,112;3,0,1)$ &(26,4) &$(16,120,192;2,2,1)$ &(27,5) &$(64,224,384;2,1,4)$    &(22,7) \\

&& && &$(16,56,224;3,0,2)$ &(27,5) &$(32,240,384;2,2,2)$ &(28,6) \\

&& && && &$(32,112,448;3,0,3)$ &(28,6)\\

&& && && &$(16,120,448;3,1,1)$ &(37,5) \\[0.2em]
\hline
\end{tabular}
\caption{Rank and kernel for all types of  $\Z_4$-linear, $\Z_8$-linear, $\Z_2\Z_4$-linear, and $\Z_2\Z_4\Z_8$-linear Hadamard codes of length $2^t$, $8\leq t\leq 11$.}
\label{table:Types1}
\end{table}

%\begin{table}[h!]
\begin{table}[!ht]
\tiny
\centering
\begin{adjustbox}{angle=90}
\begin{tabular}{|c||cc|cc|cc|cc|}
\cline{1-9}
& \multicolumn{2}{c|}{$t=12$}& \multicolumn{2}{c|}{$t=13$}& \multicolumn{2}{c|}{$t=14$}& \multicolumn{2}{c|}{$t=15$}\\
\cline{2-9}
& type & $(r,k)$ & type & $(r,k)$ & type & $(r,k)$ & type & $(r,k)$ \\[0.2em]
\hline
\multirow{2}{*}{$\Z_4$}
&$(2^{11};1,11)$ &(13,13) &$(2^{12};1,12)$ &(14,14) &$(2^{13};1,13)$ &(15,15) &$(2^{14};1,14)$ &(16,16) \\

&$(2^{11};2,9)$ &(13,13) &$(2^{12};2,10)$ &(14,14) &$(2^{13};2,11)$ &(15,15) &$(2^{14};2,12)$  &(16,16)\\

 &$(2^{11};3,7)$ &(14,11)& $(2^{12};3,8)$  & (15,12)  & $(2^{13};3,9)$  & (16,13) &$(2^{14};3,10)$ &(17,14) \\
  
&$(2^{11};4,5)$ &(16,10) &$(2^{12};4,6)$   &(17,11)   &$(2^{13};4,7)$    &(18,12) & $(2^{14};4,8)$  &(19,13)\\

&$(2^{11};5,3)$ &(19,9) &$(2^{12};5,4)$ &(20,10) &$(2^{13};5,5)$ &(21,11) &$(2^{14};5,6)$ &(22,12)\\

&$(2^{11};6,1)$ &(23,8) &$(2^{12};6,2)$  &(24,9) &$(2^{13};6,3)$ &(25,10) &$(2^{14};6,4)$ &(26,11) \\

&&  &$(2^{12};7,0)$ &(29,9) &$(2^{13};7,1)$  &(30,9) &$(2^{14};7,2)$  &(31,10) \\
&& && && &$(2^{14};8,0)$ &(37,9)\\[0.2em]

\hline
\multirow{4}{*}{$\Z_2\Z_4$} 
&$(2^{11},2^{10};1,11)$ &(13,13) &$(2^{12},2^{11};1,12)$ &(14,14) &$(2^{13},2^{12};1,13)$ &(15,15) &$(2^{14},2^{13};1,14)$ &(16,16) \\

%&$(2^{12},0;0,13)$&(13,13) &$(2^{13},0;0,14)$&(14,14) &$(2^{14},0;0,15)$&(15,15) &$(2^{15},0;0,16)$&(16,16)\\

&$(2^{10},1536;2,9)$ &(14,11) &$(2^{11},3072;2,10)$ &(15,12) &$(2^{12},6144;2,11)$ &(16,13) &$(2^{13},12288;2,12)$  &(17,14)\\

  &$(2^9,1792;3,7)$  & (16,10)& $(2^{10},3584;3,8)$  & (17,11)  & $(2^{11},7168;3,9)$  & (18,12) &$(2^{12},14336;3,10)$ &(19,13) \\
  
&$(2^8,1920;4,5)$ &(19,9) &$(2^9,3840;4,6)$          &(20,10)      &$(2^{10},7680;4,7)$          &(21,11)        &$(2^{11},15360;4,8)$  & (22,12)\\

&$(2^7,1984;5,3)$ &(23,8) &$(2^8,3968;5,4)$  &(24,9)   &$(2^9,7936;5,5)$ &(25,10) &$(2^{10},15872;5,6)$ &(26,11)\\

&$(2^6,2016;6,1)$ &(28,7) &$(2^7,4032;6,2)$  &(29,8) &$(2^8,8064;6,3)$  &(30,9) &$(2^9,16128;6,4)$  &(31,10)\\
&& && &$(2^7,8128;7,1)$ &(36,8) &$(2^8,16256;7,2)$ &(37,9)\\[0.2em]

\hline
\multirow{5}{*}{$\Z_8$}
&$(2^{10};1,0,10)$ &(13,13) &$(2^{11};1,0,11)$ &(14,14)  &$(2^{12};1,0,12)$ &(15,15)  &$(2^{13};1,0,13)$ &(16,16)\\

&$(2^{10};1,1,8)$ &(13,13) &$(2^{11};1,1,9)$ &(14,14)  &$(2^{12};1,1,10)$ &(15,15)  &$(2^{13};1,1,11)$ &(16,16)\\

&$(2^{10};1,2,6)$ &(14,11) &$(2^{11};1,2,7)$ &(15,12)  &$(2^{12};1,2,8)$ &(16,13)  &$(2^{13};1,2,9)$ &(17,14) \\

&$(2^{10};1,3,4)$ &(16,10)  &$(2^{11};1,3,5)$ &(17,11)  &$(2^{12};1,3,6)$ &(18,12)  &$(2^{13};1,3,7)$ &(19,13) \\

&$(2^{10};1,4,2)$ &(19,9) &$(2^{11};1,4,3)$ &(20,10)  &$(2^{12};1,4,4)$ &(21,11)  &$(2^{13};1,4,5)$ &(22,12) \\

&$(2^{10};2,0,7)$  &(15,10) &$(2^{11};2,0,8)$  &(16,11)  &$(2^{12};2,0,9)$  &(17,12) &$(2^{13};2,0,10)$  &(18,13)\\

&$(2^{10};2,1,5)$  &(17,9)  &$(2^{11};2,1,6)$  &(18,10)  &$(2^{12};2,1,7)$  &(19,11)    &$(2^{13};2,1,8)$  &(20,12)\\

&$(2^{10};2,2,3)$  &(20,8) &$(2^{11};2,2,4)$  &(21,9)  &$(2^{12};2,2,5)$  &(22,10)  &$(2^{13};2,2,6)$  &(23,11)\\

&$(2^{10};2,3,1)$  &(24,7) &$(2^{11};2,3,2)$  &(25,8)  &$(2^{12};2,3,3)$  &(26,9)  &$(2^{13};2,3,4)$  &(27,10)\\

 &$(2^{10};3,0,4)$  &(21,8) &$(2^{11};3,0,5)$  &(22,9)  &$(2^{12};3,0,6)$  &(23,10)  &$(2^{13};3,0,7)$  &(24,11)\\
 
 &$(2^{10};3,1,2)$  &(26,7)  &$(2^{11};3,1,3)$  &(27,8)  &$(2^{12};3,1,4)$  &(28,9) &$(2^{13};3,1,5)$  &(29,10)\\
 
 &$(2^{10};4,0,1)$  &(33,6) &$(2^{11};4,0,2)$  &(34,7)  &$(2^{12};4,0,3)$  &(35,8)  &$(2^{13};4,0,4)$  &(36,9)\\
 
 &$(2^{10};1,5,0)$ &(23,8) &$(2^{11};1,5,1)$ &(24,9)  &$(2^{12};1,5,2)$ &(25,10) &$(2^{13};1,5,3)$ &(26,11) \\
 
 &$(2^{10};3,2,0)$  &(32,6) &$(2^{11};3,2,1)$  &(33,7)  &$(2^{12};3,2,2)$  &(34,8)  &$(2^{13};3,2,3)$  &(35,9)\\
 
 && &$(2^{11};2,4,0)$  &(30,7)  &$(2^{12};2,4,1)$  &(31,8)  &$(2^{13};2,4,2)$  &(32,9)\\
 
 && &$(2^{11};4,1,0)$  &(43,6)  &$(2^{12};4,1,1)$  &(44,7)    &$(2^{13};4,1,2)$  &(45,8)\\
 
 && && &$(2^{12};1,6,0)$  &(30,9) &$(2^{13};1,6,1)$  &(31,10)\\
 
 && && &$(2^{12};3,3,0)$   &(41,7) &$(2^{13};3,3,1)$   &(42,8) \\
 
 && && &$(2^{12};5,0,0)$  &(56,6)  &$(2^{13};5,0,1)$  &(57,7)\\
 
 && && && &$(2^{13};2,5,0)$  &(38,8)\\
 
 && && && &$(2^{13};4,2,0)$  &(55,7)\\[0.2em]
\hline
\multirow{3}{*}{$\Z_2\Z_4\Z_8$} 
&$(2^{10},2^9,2^9;1,0,10)$ &(13,13) &$(2^{11},2^{10},2^{10};1,0,11)$ &(14,14)  &$(2^{12},2^{11},2^{11};1,0,12)$ &(15,15)  &$(2^{13},2^{12},2^{12};1,0,13)$ &(16,16)\\

%&$(2^{12},0,0;0,0,13)$ &(13,13)  &$(2^{13},0,0;0,0,14)$ &(14,14) &$(2^{14},0,0;0,0,15)$ &(15,15) &$(2^{15},0,0;0,0,16)$ &(16,16)\\

&$(2^9,768,2^9;1,1,8)$ &(15,10) &$(2^{10},1536,2^{10};1,1,9)$ &(16,11)  &$(2^{11},3072,2^{11};1,1,10)$ &(17,12)  &$(2^{12},6144,2^{11};1,1,11)$ &(18,13)\\

&$(2^8,896,2^9;1,2,6)$ &(18,9) &$(2^9,1792,2^{10};1,2,7)$ &(19,10)  &$(2^{10},3584,2^{11};1,2,8)$ &(20,11)  &$(2^{11},7168,2^{12};1,2,9)$ &(21,12) \\

&$(2^7,960,2^9;1,3,4)$ &(22,8)  &$(2^8,1920,2^{10};1,3,5)$ &(23,9)  &$(2^9,3840,2^{11};1,3,6)$ &(24,10)  &$(2^{10},7680,2^{12};1,3,7)$ &(25,11) \\

&$(2^6,992,2^9;1,4,2)$ &(27,7) &$(2^7,1984,2^{10};1,4,3)$ &(28,8)  &$(2^8,3968,2^{11};1,4,4)$ &(29,9)  &$(2^9,7936,2^{12};1,4,5)$ &(30,10) \\

&$(2^8,384,768;2,0,7)$  &(18,9) &$(2^9,768,1536;2,0,8)$  &(19,10)  &$(2^{10},1536,3072;2,0,9)$  &(20,11) &$(2^{11},3072,6144;2,0,10)$  &(21,12)\\

&$(2^7,448,768;2,1,5)$  &(23,8)  &$(2^8,896,1536;2,1,6)$  &(24,9)  &$(2^9,1792,3072;2,1,7)$  &(25,10)    &$(2^{10},3584,6144;2,1,8)$  &(26,11)\\

&$(2^6,480,768;2,2,3)$  &(29,7) &$(2^7,960,1536;2,2,4)$  &(30,8)  &$(2^8,1920,3072;2,2,5)$  &(31,9)  &$(2^9,3840,6144;2,2,6)$  &(32,10)\\

&$(2^5,496,768;2,3,1)$  &(36,6) &$(2^6,992,1536;2,3,2)$  &(37,7)  &$(2^7,1984,3072;2,3,3)$  &(38,8)  &$(2^8,3968,6144;2,3,4)$  &(39,9)\\

 &$(2^6,224,896;3,0,4)$  &(29,7) &$(2^7,448,1792;3,0,5)$  &(30,8)  &$(2^8,896,3584;3,0,6)$  &(31,9)  &$(2^9,1792,7168;3,0,7)$  &(32,10)\\
 
 &$(2^5,240,896;3,1,2)$  &(38,6)  &$(2^6,480,1792;3,1,3)$  &(39,7)  &$(2^7,960,3584,3,1,4)$  &(40,8) &$(2^8,1920,7168;3,1,5)$  &(41,9)\\
 
 &$(2^4,120,960;4,0,1)$  &(49,5) &$(2^5,240,1920;4,0,2)$  &(50,6)  &$(2^6,480,3840;4,0,3)$  &(51,7)  &$(2^7,960,7680;4,0,4)$  &(52,8)\\
 
&&  &$(2^6,2016,2^{10};1,5,1)$ &(34,7)  &$(2^7,4032,2^{11};1,5,2)$ &(35,8) &$(2^8,8064,2^{12};1,5,3)$ &(36,9) \\
 
&& &$(2^5,496,1792;3,2,1)$  &(49,6)  &$(2^6,992,3584;3,2,2)$  &(50,7)  &$(2^7,1984,7168;3,2,3)$  &(51,8)\\
 
 && &&  &$(2^6, 2016, 3072;2,4,1)$  &(46,7)  &$(2^7,4032,6144;2,4,2)$  &(47,8)\\
 
 && &&  &$(2^5,496,3840;4,1,1)$  &(65,6)    &$(2^6,992,7680;4,1,2)$  &(66,7)\\
 
 && && && &$(2^7, 8128, 4096 ;1,6,1)$  &(43,8)\\
 
 && && && &$(2^6, 2016, 7168 ;3,3,1)$   &(62,7) \\
 
 && && &&  &$(2^5, 496, 7936;5,0,1)$  &(85,6)\\[0.2em]
\hline
\end{tabular}
\end{adjustbox}
\caption{Rank and kernel for all types of  $\Z_4$-linear, $\Z_8$-linear, $\Z_2\Z_4$-linear, and $\Z_2\Z_4\Z_8$-linear Hadamard codes of length $2^t$, $12\leq t\leq 15$.}
\label{table:Types2}
\end{table}

\bibliographystyle{ieeetr}
\bibliography{Manuscript.bib}

\end{document}